\documentclass[useAMS,usenatbib]{mn2e}

\usepackage{amsmath}
\usepackage{amssymb}
\usepackage[dvipsnames]{xcolor}

\usepackage{graphicx}
\usepackage{graphics}
\usepackage{hyperref}
\hypersetup{
     colorlinks   = true,
     citecolor    = blue
}

\def\apj{{ ApJ}}
\def\apjl{{ApJL}}

\def\aap{{ A\&A}}

\def\mnras{{ MNRAS}}

\def\nat {{ Nature}}

\def\physrep{{ Physics Reports}}

\def\aapr{{Astro. Astrophys. Rev.}}

\def\d{\mathrm{d}}
\newcommand{\myemail}{wenbinlu@astro.as.utexas.edu}

\title[Jet-envelope interaction in TDEs]{Radiative interaction between
  the relativistic jet and optically thick envelope in tidal disruption events}
\author[Lu, Krolik, Crumley \& Kumar]
  {Wenbin Lu$^1$\thanks{\myemail}, Julian
    Krolik$^2$\thanks{jhk@jhu.edu}, Patrick Crumley$^3$,  
    Pawan Kumar$^1$ \\ 
  $^1$Department of Astronomy, University of Texas at Austin, Austin,
TX 78712, USA\\
  $^2$Physics and Astronomy Department, Johns Hopkins University,
  Baltimore, MD 21218, USA\\
  $^3$Anton Pannekoek Institute for Astronomy, University of
  Amsterdam, PO Box 94249, NL-1090 GE Amsterdam, the Netherlands}

\pagerange{\pageref{firstpage}--\pageref{lastpage}} \pubyear{2015}
\def\LaTeX{L\kern-.36em\raise.3ex\hbox{a}\kern-.15em
    T\kern-.1667em\lower.7ex\hbox{E}\kern-.125emX}

\begin{document}
\label{firstpage}
\maketitle

\begin{abstract}

Reverberation observations yielding a lag spectrum have uncovered an
Fe~K$\alpha$ fluorescence line in the tidal disruption event (TDE)
Swift~J1644+57  \citep{2016Natur.535..388K}.  The discovery paper used
the lag spectrum to argue that the source of the X-ray continuum was
located very close to the blackhole ($\sim 30$ gravitational radii)
and moved sub-relativistically. We reanalyze the lag spectrum,
pointing out that dilution effects cause it to indicate a geometric
scale an order of magnitude larger than inferred by
\citet{2016Natur.535..388K}.  If the X-ray  
continuum is produced by a relativistic jet, as suggested by the rapid
variability, high luminosity and hard spectrum, this larger scale
predicts an Fe ionization state consistent with efficient K$\alpha$
photon production. Moreover, the momentum of the jet X-rays impinging
on the surrounding accretion flow on this large scale accelerates a layer of
gas to  speeds $\sim 0.1$--$0.2c$, consistent with the blueshifted
line profile.

Implications of our results on the global picture of jetted TDEs are
discussed. A power-law $\gamma$/X-ray spectrum may be 
produced by external UV-optical photons being
repetitively inverse-Compton scattered by {\it cold} electrons in the
jet, although our model for the K$\alpha$ reverberation does not
depend on the jet radiation mechanism (magnetic reconnection in a
Poynting jet is still a viable mechanism). The non-relativistic wind
driven by jet radiation may  explain the late-time radio
rebrightening in Swift J1644+57. This energy injection may also cause the
thermal UV-optical emission from jetted TDEs to be
systematically brighter than in non-jetted ones.
\end{abstract}

\begin{keywords}
galaxies: nuclei --- accretion, accretion disks --- methods: analytical
\end{keywords}

\section{Introduction}

A new type of $\gamma/$X-ray transients---tidal disruption
events (TDEs) with jets---was established by the discovery of Swift 
J164449.3+573451 \citep[hereafter Swift J1644+57,
][]{2011Sci...333..203B, 2011Natur.476..421B, 2011Sci...333..199L,
  2011Natur.476..425Z}, Swift J2058.4+0516 \citep[hereafter Swift
J2058+05,][]{2012ApJ...753...77C} and possibly a third one Swift
J1112.2+8238 \citep{2015MNRAS.452.4297B}. They are observationally
different from other non-jetted TDEs discovered in the
UV-optical band \citep[e.g.][]{2009ApJ...698.1367G, 2012Natur.485..217G,
  2014ApJ...793...38A, 2016MNRAS.455.2918H, 2017arXiv170301299H} and
soft X-ray band
\citep[e.g.][]{2004ApJ...603L..17K, 2012A&A...541A.106S,
  2016arXiv161001788S} in that they are much brighter, have
harder X-ray spectra and vary on much shorter timescales
\citep{2011Sci...333..203B}. 

The generic model for TDEs is that they begin when a star with mass 
$M_*$ and radius $R_*$ falls toward a supermassive blackhole (BH) of
mass $M = 10^6M_6M_{\rm \odot}$ along a parabolic orbit with
pericenter distance smaller than the tidal disruption radius $R_{\rm T}$, a
distance determined by matching the BH's tidal forces to
the star's self gravity
\begin{equation}
  \label{eq:1}
  R_{\rm T} \simeq R_{\rm *} \left(\frac{M}{M_{\rm *}}\right)^{1/3},
\end{equation}
which is $\simeq (7.0 \times 10^{12}\rm\ cm)\ M_6^{1/3}
(M_*/M_\odot)^{-1/3} (R_*/R_\odot)$ for a main-sequence star.

The initial condition for a TDE is very simple, but the dynamics after
disruption are extremely complicated due to the 3 dimensional nature and
the wide range of time-/length-scales involved.  Because 
constructing a global 
deterministic model is a formidable task, various analytical and numerical
calculations have been carried out studying different aspects of the
post-disruption 
physics, e.g. fall-back stream evolution
\citep[e.g.][]{1994ApJ...422..508K, 2009MNRAS.392..332L,
  2013MNRAS.435.1809S,2013ApJ...767...25G, 2015ApJ...809..166G,
  2016MNRAS.459.3089C}, 
stream-stream collisions \citep{2016arXiv160307733J}, stream
magnetization \citep{2016arXiv160908160G, 2016arXiv161109853B}, 
disk formation \citep{2015ApJ...804...85S,
  2016MNRAS.455.2253B,2016MNRAS.458.4250S}, 
disk evolution \citep{2014ApJ...784...87S,2015MNRAS.453..157P},
possible large-scale envelope inflation \citep{1997ApJ...489..573L,
  2014ApJ...781...82C}, super-Eddington disk winds
\citep{2009MNRAS.400.2070S,2016MNRAS.461..948M},  
jet propagation \citep{2012ApJ...760..103D},
and unbound debris evolution
\citep{2016ApJ...822...48G,2016ApJ...827..127K}.  
Many of these studies are very recent and this subject is developing
rapidly.

Since different dynamical processes are deeply interconnected, it is
crucial to put together different pieces of observations to 
understand the bigger picture \citep[e.g.][]{2016ApJ...827..127K}. The three
most distinct observational components of a jetted TDE are:
non-thermal $\gamma/$X-ray emission,
thermal UV-optical emission, and non-thermal radio/mm emission.  The first is
often thought to be the result of internal dissipation within a jet; the second
is sometimes identified with the surface of the accretion flow and
sometimes with a reprocessing outflow; the third is generally
attributed to external shocks formed when an outflow runs into the
circum-nuclear medium.  In this paper, we propose 
that the radiation from the jet, upon striking the accretion flow, can drive a
kind of outflow not previously considered. 

Swift J1644+57, at redshift z = 0.354 \citep{2011Sci...333..199L}, has
a rich set of data in terms of time (minutes to years) and
multi-wavelength (radio to $\gamma$-rays) coverage; see the referenced
discovery papers. The  
non-thermal $\gamma$/X-ray emission had averaged\footnote{ 
In the first a few days, the $\gamma$/X-ray lightcurve of Swift J1644+57
had multiple hard flares, with rise time $\sim 100\rm\ s$, flare
duration $\sim 1000\rm\ s$ and quiescent interval $\sim
5\times10^4\rm\ s$. These properties have been attributed to the
disruption of a white 
dwarf (by an intermediate-mass BH) instead of a main-sequence star
\citep{2011ApJ...743..134K}. Our analysis in this paper deals with
processes happening at large radii $\gtrsim 10^{13}\rm\ cm$ and is hence
insensitive to the nature of the disrupted star and the BH mass.} isotropic 
luminosities $10^{47}$-$10^{48}\rm\ erg\ s^{-1}$ in the first 10 d
(hereafter host-galaxy rest frame time) and
then declined roughly as $t^{-5/3}$ until a sudden drop at about $370\rm\
d$ \citep{2013ApJ...767..152Z,2016ApJ...817..103M}. The X-ray
lightcurve showed very fast variability with a minimum  
variability timescale $\simeq78\rm\ s$ (host-galaxy rest
  frame time).  
The observed X-ray spectrum for Swift J1644+57 
by {\it Swift}/XRT in the 0.3(1+z)-10(1+z) keV range was a power-law
$F_{\nu}\propto \nu^{-\alpha}$ with early time (25-86 d) spectral index 
$\alpha\simeq 0.8$ and late time ($\sim$100 d) $\alpha \simeq 0.5$
\citep{2012MNRAS.422.1625S}. The early time spectrum extends up to
150(1+z) keV without a break \citep{2011Natur.476..421B}. Therefore,
most of the radiation energy is on the high frequency end, which
possibly extends to the electron rest-mass energy or higher.

Unfortunately, due to the large dust 
extinction $A_{\rm V}\sim 10\rm\ mag$, Swift J1644+57 was not observable
in the UV where the peak of the thermal disk/envelope
emission is 
predicted to be \citep[e.g.][]{1997ApJ...489..573L}. After correcting
for dust extinction, the thermal component 
can be seen in the near-infrared \citep{2016ApJ...819...51L}. Swift
J2058+05 was similar to Swift J1644+57 in the X-ray
band\footnote{Unfortunately, Swift 
J2058+05 is much farther away (by a factor of $\sim 4$) and no
reverberation observation was carried out.} but had
much less dust extinction \citep[$A_{\rm V}\sim 0.5\rm\ mag$, ][]{
2012ApJ...753...77C}. The thermal UV-optical emission from Swift
J2058+05 was observable up to 60 days after discovery 
\citep{2015ApJ...805...68P}. We call the source of the
thermal UV-optical component the {\it envelope} in general, which
could be a 
thick disk or quasi-spherical envelope inflated by radiation
pressure \citep{1997ApJ...489..573L, 2014ApJ...781...82C}, an
optically thick wind launched from the disk
\citep{2016MNRAS.461..948M}, or the shock from stream-stream
collisions \citep{2015ApJ...806..164P}. It is widely known that the
photospheric radii of the thermal UV-optical emission in TDEs are much
larger than the tidal disruption radius given by eq. (\ref{eq:1}); see
the referenced TDE discovery papers and we also provide a simple
estimate in Appendix A. We point out that, if the conversion from jet
energy to radiation occurs below the photosphere of the envelope, the
jet radiation will likely interact with and affect the dynamics of
the envelope.

An evidence of this interaction is the Fe K$\alpha$ line
detected in Swift J1644+57 by {\it XMM-Newton} at
$\sim14\rm\ d$ (and possibly also by {\it Suzaku} at $\sim 7\rm\ d$)
post-discovery \citep{2016Natur.535..388K}. The flux variations of the
Fe K$\alpha$ line in the frequency range
$(2$-$10)(1+z)\times10^{-4}\rm\ Hz$ followed the  
corresponding variations in the continuum at 4-5 keV and 8-13 keV with
a lag time of $\sim 120/(1+z) \rm\ s$. This is most naturally
explained by fluorescence of the jet X-rays off an ionized
reflector. The peak of the lag-energy spectrum is at $\simeq 8\rm\
keV$, which corresponds to a line-of-sight (LOS) velocity
$\simeq0.1$-0.2c, considering the rest line energy between 6.4 
and 6.97 keV (depending on the ionization state of Fe). Since we are 
viewing the system at a low inclination (with the jet axis close to
our LOS), such a high velocity is hard to associate with the
rotation of a disk, so the reflector must be an outflow moving away
from the disk toward us (see a schematic picture in
Fig. \ref{fig:transverse}).

In this paper, we study the interaction between jet radiation
and the surrounding envelope and discuss how this interaction affects
the dynamics 
of the envelope and implications on the global picture of jetted TDEs. In 
section \ref{sec:jet_motion}, we discuss the physical constraints on the
jet Lorentz factor and show that the jet is moving at a relativistic
speed. In section \ref{sec:envelope}, we discuss the physical state of
the envelope based on observations of Swift J2058+05, a close analog
to Swift J1644+57. In section \ref{sec:Kalpha} and \ref{sec:lag}, we
study how the observed K$\alpha$ line is produced in the jet-envelope
interaction. Implications of our results and some possible issues are
discussed in section \ref{sec:discussion}. A short summary is in
section \ref{sec:summary}. Unless otherwise
clearly stated, all frequencies, time and
luminosities in this paper have been de-redshifted to the
host-galaxy rest frame (for Swift J1644+57 at $z = 0.354$). Throughout
the paper, the convention $Q = 10^nQ_n$ and CGS units are used.

\section{Relativistic motion of the jet}\label{sec:jet_motion}
In this section, we derive a conservative lower limit of the bulk
Lorentz factor of the $\gamma$/X-ray emitting plasma from simple
Compton scattering arguments. The radiation seen by
the observer is also impinging and exerting 
a Compton force on electrons in the (optically thin) source. This is
equivalent to the case where an electron is at a distance $R$ from a
point source of the same luminosity.

Let this electron move with speed $\beta$ (Lorentz factor
$\Gamma=(1-\beta^2)^{-1/2}$) radially outward from an isotropic
radiation source of $L$.
For a baryonic jet, the electron, with the inertia of a proton
$m_{\rm p}$, experiences an acceleration from Compton scattering
\begin{equation}
  \label{eq:16}
{\d p'\over \d t'} = \Gamma^2(1- \beta)^2 
  \frac{L\sigma_{\rm T}}{4\pi R^2 c},
\end{equation}
where the momentum and time in the electron's comoving frame are
denoted with a 
prime ($'$) and the R.H.S. is the momentum flux in the comoving frame
multiplied  by the Thomson cross section $\sigma_{\rm T}$. Going back
to the rest frame of the BH, it can be shown that $\d p'/\d t' =
m_{\rm p} c\ \d(\Gamma\beta)/\d t$.

Using $\mathrm{d} R = \beta c \mathrm{d} t$, $R = 10^{13}R_{13}\rm\
cm$ and $L = 10^{47}L_{47}\rm\ erg\ s^{-1}$, we obtain
\begin{equation}
  \label{eq:17}
  \frac{\Gamma\beta \mathrm{d} \beta}{(1-\beta)^2} = 11.8
  L_{\rm 47} \frac{\mathrm{d} R_{13}}{R_{13}^2},
\end{equation}
which can be integrated analytically. For initial condition
$\beta=0$, an electron accelerated from $R$
to $2R$ attains a Lorentz factor given by
\begin{equation}
  \label{eq:38}
 \frac{2\beta - 1}{3\Gamma(1-\beta)^2} = 5.9 
  \frac{L_{\rm 47} }{R_{13}} -\frac{1}{3}.
\end{equation}
With bolometric correction of a factor of a few, the typical
$\gamma$/X-ray luminosity of Swift J1644+57 within the first $\sim 10 $ 
days is $L \simeq 3\times 10^{47}\rm\ erg\ s^{-1}$, so an electron
accelerated from $R$ to $2R$ attains a Lorentz factor of $\Gamma
= (2.6, 1.9, 1.5, 1.3)$ when $R = (1, 3, 10, 20)\times10^{13}\rm\
cm$. These are conservative lower limits for the jet Lorentz factor at
the radius where $\gamma$/X-rays are produced.

The X-ray emission from Swift J1644+57 
had a minimum variability timescale $t_{\rm var, min} = 78\rm\ s$. If the X-ray
source is moving at Lorentz factor $\Gamma$ toward the
Earth and the comoving size of the emitting region is $R/\Gamma$ (the
causally connected region), the jet radiation radius can be estimated
by \citep{2011Sci...333..203B} 
\begin{equation}
  \label{eq:11}
  R \simeq \Gamma^2c t_{\rm var, min} = 2.3\times
10^{12}\Gamma^2\rm\ cm.
\end{equation}
Combining eqs.  (\ref{eq:38}) and (\ref{eq:11}), one obtains $\Gamma
\simeq 
2.4$, which may be considered as a lower limit of the Lorentz factor
at the radius where the jet is radiating.

Note that
the variability timescale could be affected by many factors (e.g. the
comoving size of the emitting region could be much smaller than $
R/\Gamma$), so we do not use $t_{\rm var, min}$ as a hard constraint
on the relation between $R$ and $\Gamma$. We also note that, for a
baryonic jet with hot electrons (electrons' Lorentz factors in the
comoving frame $\gamma_{\rm e}>1$) or with lepton-to-proton number
ratio larger than 1, the lower limit on $\Gamma$ will be stronger.

On the other hand, if the jet is Poynting-dominated and the
lepton-loading is very low, the effective inertia of an 
electron could be larger than $m_{\rm p}$, so Compton
acceleration may not be efficient\footnote{Poynting-dominated jets
  could be dissipative over certain range of distances from the
  central engine. The following two reasons could lead to a larger
  acceleration: (1) an outward gradient of 
  magnetic pressure when dissipation causes magnetic energy to
  decrease with distance \citep[e.g.][]{2002A&A...391.1141D}; (2)
  the Compton rocket effect when leptons are 
  accelerated to high Lorentz factors in the comoving frame
  \citep[e.g.][]{1981ApJ...243L.147O,1982MNRAS.198.1109P}.}. However,
the jet will accelerate as 
a result of its own magnetic pressure gradient. The fast X-ray
variability implies that the jet is not continuous but intermittent
with individual ``blobs'' likely having durations $t_0 \lesssim t_{\rm
  var,min}$. If the initial magnetization of such a blob
$\sigma_0\gg 1$, it quickly accelerates to a Lorentz factor
$\Gamma\simeq \sigma_0^{1/3}$ after propagating a distance of $R_0\sim 
ct_0\lesssim 2.3\times10^{12}\rm\ cm$, and then the Lorentz factor
increases with radius $R$ as $\Gamma\sim (\sigma_0 R/R_0)^{1/3}$ until
the saturation radius $R_{\rm s}\simeq \sigma_0^2R_0$
\citep{2011MNRAS.411.1323G}. For instance, for $\sigma_0\simeq 10$,
the jet in Swift J1644+57 can accelerate to $\Gamma\gtrsim 5$ at $\sim
10^{13}\rm\ cm$ from the BH. The acceleration could be even faster
when there is extra collimation. 
\citet{2007MNRAS.380...51K} show that for $\sigma_0\simeq 10$,
Poynting-dominated jets accelerate to $\Gamma\simeq 5$ at
$10$-$30R_{\rm lc}$, where $R_{\rm lc}\simeq 4 R_{\rm g}$ is the light
cylinder radius of a fast spinning ($a\simeq1$) Kerr BH and $R_{\rm g} =
GM/c^2$ is the gravitational radius.

We also note that the early time $\gamma$/X-ray spectrum of Swift
J1644+57 is a power-law 
$F_{\nu}\propto \nu^{-\alpha}$ ($\alpha\simeq 0.8$) extending up to
200 keV without a break \citep{2011Natur.476..421B,
  2011Sci...333..203B}. If this power-law continues extending up by a
factor of a 
few to $\sim$ MeV energy, relativistic motion is 
needed to alleviate the compactness problem. For example, for
luminosity $\nu L_{\nu}|_{\rm MeV}\simeq 10^{47}\rm\ erg\ s^{-1}$
and minimum variability time $t_{\rm var, min} \simeq 10^2\rm\ s$, the optical
depth for pair production reaches $\tau_{\gamma\gamma}\sim 10^5$,
which can be reduced by a factor of $\Gamma^{-6+2\alpha}$ if the
source is moving toward the Earth at Lorentz factor $\Gamma$.

Another argument for a relativistic jet is that, when the X-ray source
is non-relativistic, Fe ions are fully stripped and hence K$\alpha$
production is strongly suppressed (see section \ref{sec:Kalpha}).

\section{Physical state of the envelope}\label{sec:envelope}
In this section, we discuss the physical properties of the thermal
UV-optical source. We call it the
{\it envelope} because it wraps around the relativistic
jet. This region is the presumptive source of K$\alpha$ photons. We
discuss its temperature, photospheric radius, density, pressure, etc.
Interaction between jet radiation and the envelope 
will be discussed in the next section.

When the bound stellar mass returns to the vicinity of the BH, several
shock systems are formed \citep{2015ApJ...804...85S}.  One, the
``nozzle shock", 
forms near the stellar pericenter as debris streams converge toward
the orbital plane; 
this shock deflects a minority of the bound mass in toward the BH, while
a majority of the material continues along the highly-elliptical orbits.
Other shocks form where the streams intersect in the orbital plane.
If the stellar pericenter $R_{\rm p}\gtrsim 10 R_{\rm g}$ (where
$R_{\rm g}\equiv GM/c^2$), these shocks are near the apocenters of the
stream  orbits, ultimately creating an extended accretion flow on the
scale of the semi-major axis of the most-bound debris, $\sim
(M/M_*)^{1/3} R_{\rm T}$.  On the other hand, if the 
stellar pericenter $R_{\rm p} \lesssim 10R_{\rm g}$, relativistic apsidal precession is strong
enough to bring those shocks in to radii not much greater than $R_{\rm
  p}$ \citep{2015ApJ...812L...39D}.
Because the in-plane shocks dissipate an amount of energy comparable to the
binding energy of the orbit, and the cooling time is generally comparable to
or longer than the orbital timescale \citep{2015ApJ...806..164P}, the resulting
structure is geometrically thick.  However, because some of its support remains
rotational, a cone aligned with the angular momentum axis is left almost clear
of gas (this picture is complicated by Lense-Thirring dynamics, a point we
neglect for the time being).  If our viewing angle lies within the opening
angle of the cone, the likely state when we are
seeing radiation from a jet, we can see the inner surface of this cone
directly. 

The radial scale on which the majority of the mass is distributed can be
estimated in two ways.  One is the semi-major axis scale of the most-bound
debris already mentioned.  Another is phenomenological, by identifying the
UV-optical emission with thermal radiation from the photosphere of the
optically thick gas and measuring (or at least constraining) its
temperature from the observed spectrum.  This 
latter method may have large uncertainty when applied to Swift~J1644+57
because of its large extinction (we provide a rough estimate in
  Appendix A). On the other hand, 
if we suppose that other TDEs can be used as stand-ins for
Swift~J1644+57, it can be applied to 
them. In the case of many ordinary ``thermal" non-jetted TDEs, this
method yields radial scales quite consistent with the semi-major axis
estimate \citep{2015ApJ...806..164P}; see also Appendix A for simple
estimates.

Perhaps a closer analog to Swift~J1644+57, however, is Swift~J2058+05,
whose output was similarly dominated by hard X-rays.  In this case,
UV-optical continuum 
was detected, but the thermal component was entirely 
in the Rayleigh-Jeans (R-J) limit, and could therefore place only a
lower bound on the 
temperature $kT>h\nu_0$ ($\nu_0$ being the detection frequency). We may
place a upper bound on the radiating area of Swift~J2058+05 because the
specific luminosity from a thermal surface in the R-J limit is $L_{\rm
  \nu_0}\propto R_{\rm ph}^2\nu_0^2T$. For Swift~J2058+05, the
specific luminosity at $\nu_0 = 1.0\times10^{15}$ Hz is
$\nu_0L_{\nu_0} = 1.3\times10^{44}\rm\ erg\ s^{-1}$
\citep{2015ApJ...805...68P} about $11$ days post discovery, and then this
argument implies a radial scale at most $\sim 5 \times
10^{14}$~cm. On the other hand, the bolometric luminosity from a
thermal surface is $L_{\rm bol}\propto R_{\rm ph}^2T^4$ and it should
be less than $\sim 10^{47}\rm\ erg\ s^{-1}$ given by the total energy
budget\footnote{About half of
  the star remains bound and about half of the bound mass falls back
  within the Keplerian period of the most bound orbit. Other
  processes, e.g. adiabatic expansion and accretion efficiency being
  less than unity, will likely reduce the radiation energy to $\sim
  10\%$ of the total rest mass energy available.} of $\sim 10^{53}\rm\ erg$
and duration $\sim 10^6\rm\ s$. This argument gives a radial scale at least
$\sim 2\times 10^{14}$~cm. Although these arguments are based on
spherical-symmetry assumption, we do get a good insight on the radial
scale of the thermal envelope being a few$\times10^{14}$~cm.

This radial scale estimate describes Swift~J2058+05, but we will
henceforward adopt $10^{14}$~cm as a fiducial radial scale for the accretion 
flow around Swift~J1644+57.  In terms of this fiducial scale, we can make several
estimates defining characteristic conditions of the bound gas of total
mass $M_{\rm b}$.  Assuming solar metallicity with Thomson opacity
$\kappa_{\rm T}=0.34\rm\ cm^2\ g^{-1}$, we have the Thomson optical
depth $\tau_T \simeq 2 \times 10^{3} (M_{\rm b}/0.4M_\odot)
R_{14}^{-2}$ and mean mass density 
\begin{equation}
  \label{eq:9}
  \bar \rho \simeq (2 \times
10^{-10}\rm\ g\ cm^{-3})\ (M_{\rm b}/0.4M_\odot) 
R_{14}^{-3},
\end{equation}
implying an electron number density of $\bar{n}_{\rm e}\simeq (1
\times 10^{14}\rm\ cm^{-3})\ (M_{\rm b}/0.4M_\odot) R_{14}^{-3}$. For
photons with energy near 
$kT$, the bound-free opacity, which dominates over free-free opacity
in the temperature and density ranges of interest, is $\kappa_{\rm
  bf}\simeq (0.4 \rm\ cm^2\ g^{-1})\ T_5^{-7/2}(M_{\rm b}/0.4M_\odot)
R_{14}^{-3}$, which is comparable to the Thomson opacity $\kappa_{\rm
  T}$. We scale here to a temperature of $10^5$~K to 
be consistent with the R-J character of the Swift~J2058+05 UV-optical
spectrum \citep[$T\gtrsim 6\times10^{4}\rm\
K$,][]{2012ApJ...753...77C}. Also for Swift J2058+05, 
the energy budget constraint $L_{\rm 
  bol}\lesssim10^{47}\rm\ erg\ s^{-1}$ restricts the temperature to be
$T\lesssim2.5\times10^5\rm\ K$.

\begin{figure}
  \centering
\includegraphics[width = 0.3 \textwidth,
  height=0.31\textheight]{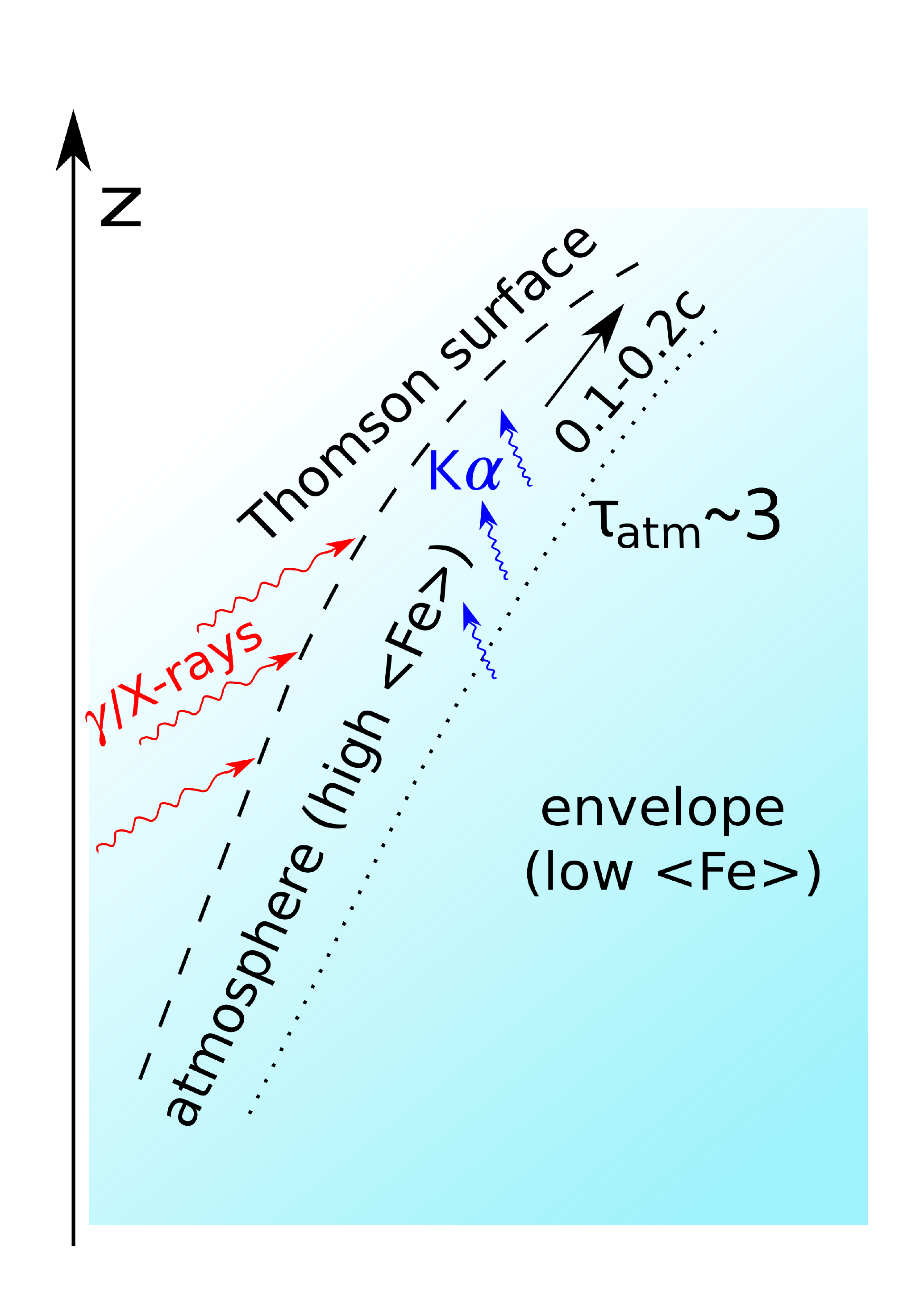}
\caption{Schematic picture of the transition region between the jet and the
  optically thick envelope. The $\gamma$/X-rays from the jet (in red
  wiggly lines) impinge on the optically thick envelope (light blue
  region). The surface where the Thomson depth of the envelope $\tau =
  1$ is called the {\it Thomson surface} and denoted by a black dashed
  curve. The part of the envelope with Thomson depth $\tau_{\rm
    atm}\sim 3$ where the $\gamma/$X-rays can diffuse
  to is called the {\it atmosphere}. The mean value of the Fe
  ionization state $<Fe>$, 
  i.e. the averaged number of positive charges per Fe ion, decreases
  as we go deeper into the envelope. We show in
  \S 4.1 that the conditions such that Fe ions have a few
  bound electrons can be realized in the
  atmosphere, so K$\alpha$ line photons
  (in blue wiggly lines) can be naturally produced in those regions.
  We also show in \S 4.3 that the atmosphere can be
  accelerated to speeds $\sim 0.1$-$0.2c$ by the momentum kick
  from the incident $\gamma/$X-rays.
}\label{fig:transverse}
\end{figure}

In these conditions, radiation pressure $aT^4/3 \simeq 2 \times
10^5 T_5^4\rm\ dyne\ cm^{-2}$ ($a$ being the radiation
constant) dominates over gas pressure
$1.4\bar{\rho} kT/m_{\rm p} \simeq (2 \times 10^3 \rm\ dyne\ cm^{-2})\
T_5(M_{\rm b}/0.4M_\odot) R_{14}^{-3}$. If the isotropic equivalent X-ray
luminosity $L_{\rm X} \sim 10^{47}\rm\
erg\ s^{-1}$ were aimed along the normal to
the inner surface of the gas cone, it would exert a pressure $\sim 3
\times 10^7 R_{14}^{-2}\rm\ dyne\ cm^{-2}$, much greater 
than either the gas or thermal radiation pressure we have
estimated. However, if the jet is both relativistic and runs along the
axis of the open cone, the flux 
striking the surrounding gas is reduced relative to its maximally beamed
value by both relativistic beaming and geometric projection.  For a
relativistic point source with
spectrum $F_\nu \propto \nu^{-\alpha}$ \citep[$\alpha \simeq
0.8$ in the first a few weeks, ][]{2011Natur.476..421B,
  2011Sci...333..203B}, the reduction due to the former is $\simeq
[\Gamma^2 (1 - \beta\cos\theta)]^{-(3+\alpha)}$ in the 
polar-angle direction $\theta$ from a jet 
traveling at $\beta c$ and Lorentz factor $\Gamma \gg 1$; the latter
is $\cos\psi$, where $\psi$ is the angle between the ray direction 
from the source to the surface and the surface normal. It is
also possible that the flux incident on the surrounding gas can be
augmented at order-unity level by X-rays that are reflected by the
one part on the inner cone surface and then strike again somewhere
else across 
the cone. The reduction in pressure is another factor of $\cos\psi$
times the reduction in flux. The location of the cone surface may  
be determined by pressure balance from the jet radiation,
centrifugal force, gravity and the intrinsic gas thermal pressure. A
schematic picture of the transition region between the jet and the
optically thick envelope is shown in Fig. (\ref{fig:transverse}).

\section{Iron K$\alpha$ emission line}\label{sec:Kalpha}
Knowing the properties of the surrounding envelope, we discuss the
conditions required to produce a K$\alpha$ line consistent with
observations. We focus on the ionization state of Fe (\S 4.1) and
equivalent width (\S 4.2) and broadening (\S 4.3) of the K$\alpha$
line. At the end (\S 4.4), we propose that the momentum of the jet
X-rays impinging on the envelope accelerates a layer of gas to speeds
$\sim$0.1-0.2$c$, consistent with the observed blueshift. Lag time will be
discussed in section \ref{sec:lag}.

\subsection{Ionization state of Fe}
\citet{2016Natur.535..388K} reported a detection of an Fe~K$\alpha$
emission line in the lag-energy spectrum from Swift~J1644+57 and
explained it 
as due to fluorescence from an outflowing funnel wall exposed to
X-rays from a sub-relativistically moving source close to the BH.
The lag-energy spectrum related variations of the line
flux and the continuum over the frequency range (2-$10)(1+z)\times10^{-4}$ Hz, and
the observed peak lag was $\sim 120/(1+z)\rm\ s$.

In their
kinematical toy model, the X-ray source is stationary at a point on 
the funnel axis $30 R_{\rm g}$ from the BH, the funnel opening angle is
$30$-$45^{\rm o}$, and the gas at the funnel wall accelerates radially
outward from $0.1c$ to $\sim 0.5c$.  The wall fluoresces from $6
R_{\rm g}$ out to $200 R_{\rm g}$. They also suggest a BH mass of $2
\times 10^6 M_\odot$, which places the X-ray source at $9 \times
10^{12}$~cm from the BH.

However, this model is problematic in several ways. The first has
to do with the ionization state of the gas at the funnel wall.  Since
the X-rays are emitted isotropically in the model, the luminosity
per unit solid angle incident on the funnel is the same as we see,
$L_{\rm X}=4\pi \d L_{\rm X}/\d\Omega = 10^{47}L_{\rm X,47}$~erg~s$^{-1}$ in the
0.3(1+z)-10(1+z) keV band.
If we focus on ionization of H-like Fe (Fe$^{25+}$), the threshold
energy is $h\nu_{\rm th}=9.3$~keV,
so only the portion of the spectrum above that energy is relevant.  The
observed spectral slope $L_\nu \propto \nu^{-0.8}$ leads to a
luminosity $\nu L_{\nu}|_{\nu_{\rm th}} = 0.37L_X$ at the threshold
energy. With the assumption that the distance from the 
BH to the X-ray source ($\sim$$10^{13}$~cm) is the same as the
characteristic distance from the source to the funnel (a good
approximation for this conical geometry), the X-ray flux incident on
the funnel is $\nu F_{\nu}|_{\nu_{\rm th}}\simeq 3 \times
10^{19}$~erg~cm$^{-2}$. Then, the photoionization timescale for Fe$^{25+}$
is
\begin{equation}
  \label{eq:30}
  \begin{split}
      t_{\rm pi} &= \left(\int_{\nu_{\rm th}}^{\infty} \sigma_{\rm
      pi}(\nu) \frac{F_{\nu}}{h \nu} \mathrm{d} \nu \right)^{-1} =
  \frac{ (3+\alpha) h\nu_{\rm th}}{\sigma_{\rm pi,th}(\nu
    F_{\nu}|_{\nu_{\rm th}})}\\ 
  &\simeq (6 \times10^{-8}\hbox{~s})\ L_{\rm\ X,47}^{-1}\\
  \end{split}
\end{equation}
where we have used the photoionization cross section
$\sigma(\nu) = \sigma_{\rm pi,th} (\nu/\nu_{\rm th})^{-3}$ and
$\sigma_{\rm pi,th}\simeq 3.3\times10^{-20}\rm\ cm^2$ for Fe$^{25+}$
\citep{1991MNRAS.249..352G}.
On the other hand, the recombination timescale
from Fe$^{26+}$ to Fe$^{25+}$ is 
\begin{equation}
  \label{eq:31}
  t_{\rm rec} \simeq \left(\alpha_{\rm A} n_{\rm e}\right)^{-1}\simeq
  (5 \times 10^{-5}\hbox{~s})\ (\rho/{\bar \rho})^{-1} T_{5}^{0.7},
\end{equation}
where we have used the Case A recombination rate coefficient
$\alpha_{\rm A}\simeq (2.0 \times 10^{-10} \mathrm{~cm^3~s^{-1}})\
T_5^{-0.7}$  
\citep{2011piim.book.....D} 
and the mean density $\bar{\rho}$ estimated in eq. (\ref{eq:9}). Comparing
eq.~(\ref{eq:30}) with eq.~(\ref{eq:31}), 
we see that only $\sim 10^{-3}$ of all Fe ions retain even a single
electron, and the K$\alpha$ production rate is
correspondingly suppressed. Note that we previously justified an
estimated temperature for the funnel wall of $\sim 10^5$~K on the
grounds that it had to be hot enough to make the UV-optical continuum
of Swift~J2058+05 entirely within the Rayleigh-Jeans range; strong
photoionization can drive the temperature to at least this level. 

Second, from section \ref{sec:jet_motion}, we know that a baryonic
dominated source will be accelerated by the radiation pressure to
Lorentz factors $\Gamma\gtrsim 2.5$ for if the $\gamma$/X-rays are
produced at $R\lesssim 10^{13}\rm\ cm$. A magnetic energy dominated
source will also accelerate to relativistic speeds due to its own
magnetic pressure gradient. Moreover, as we have shown in section
\ref{sec:envelope}, for a sub-relativistic source from which
$\gamma/$X-rays are emitted nearly isotropically, the radiation
pressure of the $\gamma/$X-rays will push the envelope out to much
greater distance, vitiating the lag-time argument in
\citet{2016Natur.535..388K}.

Third, it is unclear how a sub-relativistic source can produce a hard
$\gamma$/X-ray spectrum \citep{2011Natur.476..421B,
  2011Sci...333..203B}, given that the  
source is suffused by a large injection of thermal seed photons from
the surrounding envelope.

All the problems in the model of \citet{2016Natur.535..388K} can be
solved readily by two modifications: (i) moving the radial scale of
the fluorescing matter outward by an order of magnitude, to the scale
suggested by tidal disruption dynamics, and (ii)
putting the source of $\gamma/$X-rays in a relativistic jet.  We will
reconcile the latter change with the lag 
spectrum in section ~\ref{sec:lag}.

Moving the fluorescing matter to a radius 
10 times greater ($\sim$$10^{14}$~cm) reduces the flux by a factor of
$\sim 10^{-2}\cos\psi$, where $\psi\gtrsim 60^{\rm o}$ is the angle
between X-ray direction and the surface normal (see Fig. \ref{fig:EW}). If the
$\gamma/$X-ray source is, in fact, a relativistic jet, then the flux
on the envelope surface can be reduced 
by an additional factor of $\eta_{\rm rel} \lesssim 0.1/(\cos\psi)$ by
placing the surface outside the  
relativistic beaming cone. As shown later in \S 4.2, this
reduction factor is related to the K$\alpha$ equivalent width and is
consistent with observations. Combining these two effects, the
photoionization 
timescale can easily rise to be comparable to or greater than the
recombination time, so that most Fe ions retain at least 
one electron. As already remarked, the same reduction in flux due to jet-beaming
also permits an equilibrium between the X-ray radiation pressure and the
envelope pressure.  Lastly, bulk Comptonization can lift the thermal
UV-optical photons from the surrounding envelope to the X-ray band
without 
requiring relativistic electrons in the comoving frame of the source.

In a more exact treatment, it would be necessary to consider more carefully
the finite-thickness transition layer between the bulk of the surrounding
gas and the low-density region within the cone.  In that transition layer,
the density declines below the mean, lengthening the recombination time.
Determining the sharpness of this density cut-off is beyond the scope
of the present work.  We only comment that the density profile is
probably not exponential because it is not determined by balancing a
pressure gradient with gravity; instead, the principal mechanism is
penetration of  $\gamma$/X-rays into the gas layer and 
consequent heating.  For this reason, its characteristic thickness is likely
determined by the mean free path for Compton scattering. Thus, under
these circumstances, the fluorescence efficiency should be fairly
high.
\subsection{K$\alpha$ equivalent width}

\begin{figure}
  \centering
\includegraphics[width = 0.25 \textwidth,
  height=0.28\textheight]{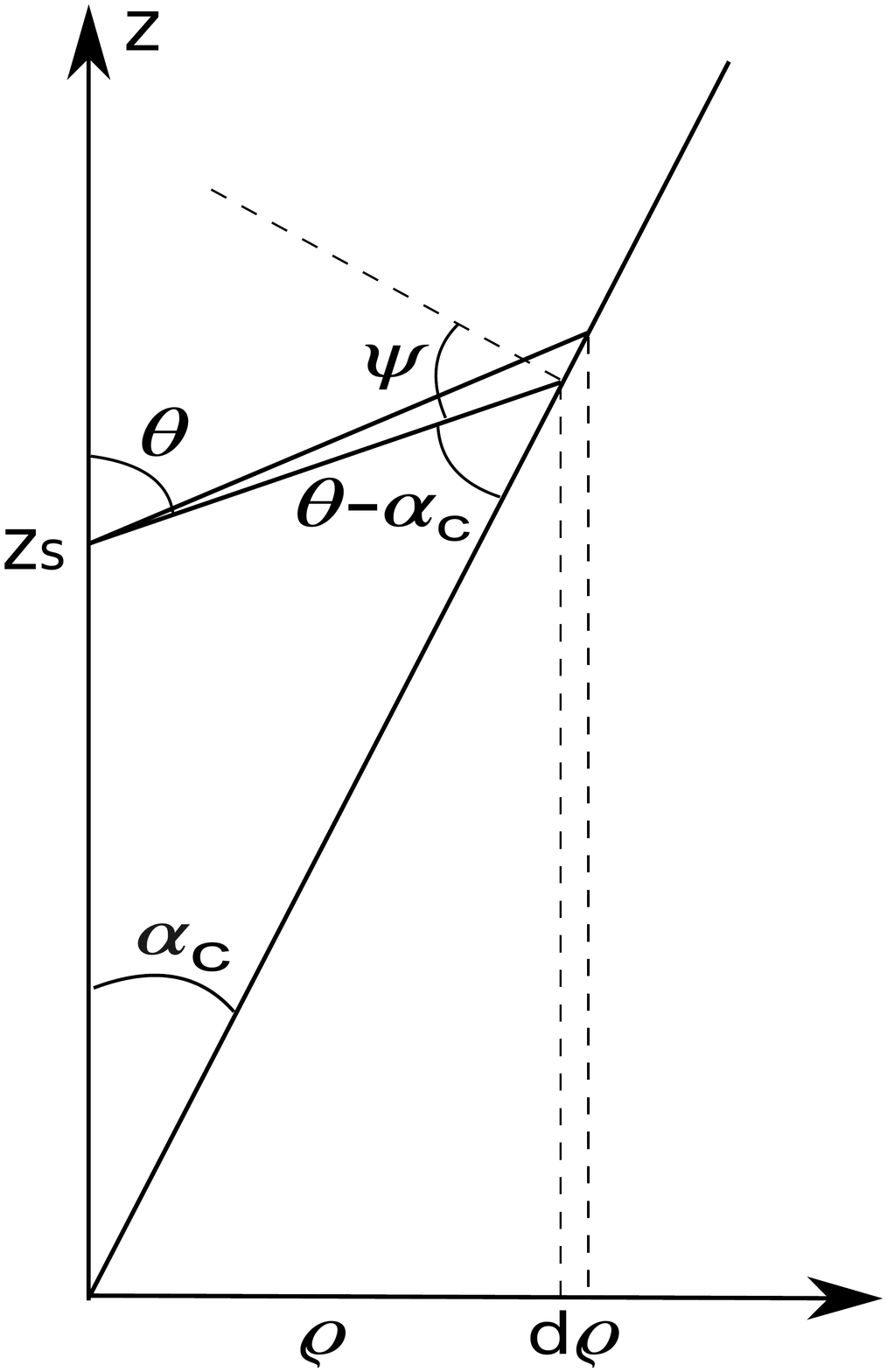}
\caption{Conical inner surface geometry. The X-ray source is on the z-axis at
  $z_{\rm s}$. The polar angle of the inner surface of the envelope is
  $\alpha_{\rm c}$. We denote the polar angle of light rays from the
  source as $\theta$ and the cylindrical radius of the illuminated
  point on the inner surface as $\varrho$. The relation between $z_{\rm s}$
  and $\varrho$ is given by $\cot \theta = \cot\alpha_{\rm c} - z_{\rm s}/\varrho$. The
  angle between light rays from the source and the surface normal is
  $\psi = \pi/2-(\theta - \alpha_{\rm c})$.
}\label{fig:EW}
\end{figure}

Given a geometry for the envelope's inner surface, the equivalent width (EW)
of the line can be readily calculated.  For example, consider a simple
model shown in Fig. (\ref{fig:EW}) in which the inner surface is
exactly conical, with half-opening angle $\alpha_{\rm c}$, and we have
\begin{equation}
\label{eq:equiv_width}
\begin{split}
  EW &= 2\epsilon_{\rm K\alpha}Y \int \frac{\d \varrho}{\varrho}\eta_{\rm
  rel}(\theta)\sin^2\theta
\sin(\theta-\alpha_{\rm c})
\int_{\frac{\epsilon_{\rm K}}{\epsilon_{\rm K\alpha}}} \frac{\d x}{x^{1+\alpha}} 
f(\epsilon), \\
&\simeq 2\epsilon_{\rm K\alpha} Yf(\epsilon_{\rm K}) \int \frac{\d
  \varrho}{\varrho}\eta_{\rm rel}(\theta)\sin^2\theta
\sin(\theta-\alpha_{\rm c})
\end{split}
\end{equation}
where $\theta(\varrho)$ is the polar angle of light rays from the
source, $\varrho$ 
is the cylindrical radius of the illuminated point on the inner
surface of the envelope, $\eta_{\rm rel}[\theta(\varrho)]\equiv
[\d L/\d\Omega (\theta)]/(\d
L/\d\Omega|_{\rm LOS})$ is the ratio  
between the intensity striking on the surface and the intensity beamed
along the LOS (which we assume to be along the jet
axis), $\epsilon_K$ and $\epsilon_{K\alpha}$ are the energies of 
K-edge and K$\alpha$ photons respectively, the dimensionless
integration variable $x \equiv \epsilon/\epsilon_{\rm K\alpha}$, 
$f(\epsilon)$ is the fraction of incident X-rays of energy $\epsilon$ that
photoionize an Fe ion before escaping, $Y$ is
the fluorescent yield. In the second row of
eq. (\ref{eq:equiv_width}), we have used $\int_{\epsilon_{\rm
      K}/\epsilon_{\rm K\alpha}} f(\epsilon) x^{-1-\alpha}\d x\simeq
 f(\epsilon_{\rm K})$, for the continuum power-law index $\alpha \simeq
 0.8$, $\epsilon_{\rm K}/\epsilon_{\rm K\alpha}\simeq 4/3$ and
 $f(\epsilon)$ being a relatively smooth function.

Then we put numbers into eq. (\ref{eq:equiv_width}): $\epsilon_{\rm
  K\alpha}\simeq 7\rm\ keV$, $f(\epsilon_{\rm K})\sim
    1$\footnote{For solar metallicity ($n_{\rm Fe}\simeq
    3.5\times10^{-5}n_{\rm 
    H}$), the K-edge absorption
  opacity $\kappa_{\rm K}$ nearly equals to the Thomson opacity
  $\kappa_{\rm T}$, if all Fe ions are H-like ($\kappa_{\rm K}$
  is larger for less ionized Fe). Compton scattering does not destroy
  photons. In the narrow jet-cone geometry of Swift J1644+57, a photon
  reflected from the funnel wall will most likely hit the
  other side of the inner surface and eventually gets absorbed.}, $Y\simeq 0.6$ 
\citep[atomic yield for Fe$^{23+, 24+, 25+}$,
][]{1987ApJ...320L...5K}, and $\d \varrho/\varrho\sim 1$. 
The trigonometry factor $\sin^2 \theta \sin (\theta -
\alpha)$ is roughly 0.05 with uncertainty\footnote{
For ($\theta = 20^{\rm o}, \alpha = 10^{\rm o}$), we have $\sin^2 \theta
\sin (\theta -\alpha)\simeq 0.02$; for ($\theta = 30^{\rm o}, \alpha =
20^{\rm o}$), it is $\simeq 0.04$; for ($\theta = 30^{\rm o}, \alpha =
10^{\rm o}$), it is $\simeq 0.09$; for ($\theta = 45^{\rm o}, \alpha =
30^{\rm o}$), it is $\simeq 0.13$.} of a factor of $\lesssim
3$. Therefore, we obtain $EW \sim 400 \bar{\eta}_{\rm rel}\rm\
eV$, where $\bar{\eta}_{\rm rel}$ is
the X-ray intensity correction due to relativistic
beaming averaged over the reprocessing surface.  In
eq. (\ref{eq:equiv_width}), we have only included the K$\alpha$   
photons emitted directly from the reprocessing surface. If the opening
angle of the funnel wall $\alpha_{\rm c}\lesssim 30^{\rm o}$, a large fraction of
K$\alpha$ photons could bounce between the inner surface of the wall
before eventually escaping. These reflected K$\alpha$ photons
  have lower energies and longer lags with respect to continuum
  variations than the photons that escape directly without reflection, and they
  produce the red wing in the lag-energy spectrum (see discussion
  later in \S 4.3).

Modeling the detailed production and
radiative transfer of K$\alpha$ photons is out of the scope of the
current work. Considering the purpose of this paper being
understanding the broad-brush picture of the jet-envelope interaction
in Swift J1644+57, we hereafter use the rough estimate $EW \sim 400
\bar{\eta}_{\rm rel}\rm\ eV$, which has uncertainty up to a factor of 3 mostly
due to the unknown geometry of the system.

We argued in \S 4.1 that in order for most Fe atoms to be H-like or
less-ionized, the intensity ratio due to relativistic beaming
$\bar{\eta}_{\rm rel} \lesssim 0.1/(\cos \psi)$, where 
$\psi = \pi/2 - (\theta - \alpha_{\rm c})$ may vary in the range
(60$^{\rm o}$, 90$^{\rm o}$) and hence
$\cos\psi\lesssim 0.5$.
This means that the EW of the K$\alpha$ line in Swift~J1644+57 should
be $\lesssim 80$~eV (note that a 
modest ratio $\bar{\eta}_{\rm rel}\sim 0.2$ implies that typical values 
for $\theta(\varrho)$ are at most a few times $1/\Gamma$).
In the time-integrated flux spectrum from {\it XMM-Newton}, there is a narrow
peak at 8~keV (host-galaxy rest frame; this is roughly the same energy as the
peak in the lag-energy spectrum) 
with an EW of $60\pm 10\rm\ eV$ \citep{2016Natur.535..388K}, which
agrees with our model.

However, this apparent agreement should be
viewed cautiously.  In {\it Suzaku} data taken 7 days earlier, the EW is 
$<7$ eV \citep{2016Natur.535..388K}, suggesting that the K$\alpha$ EW is
subject to sizable fluctuations, perhaps caused by the system's
changing geometry or Fe ionization state. As we will discuss in
section~\ref{sec:lag}, the K$\alpha$ EW plays an
important role in interpreting the lag spectrum. Due to possible
fluctuations, we do not take $EW\simeq 60\rm\ eV$ 
  as a hard constraint on the system's parameters, in order to be
  conservative. Future multi-epoch
observations of more jetted TDEs may confirm or falsify the existence
of such a narrow K$\alpha$ line.

If it is true, this narrow line gives interesting constraints on
  the differential velocity, electron temperature and optical depth of
  the line emitting region. (1) The full-width at half maximum (FWHM) in the flux
  spectrum is only $\sim 0.2$~keV, which, combined with the blueshift velocity 
$v\sim 0.1$-$0.2c$, means the velocity difference between different parts
of the line emitting gas is only $\delta v/v\sim 20\%$. This means
these K$\alpha$ photons are produced in a area where the gas has
nearly uniform LOS velocity. (2) Compton scattering broadens the
K$\alpha$ line due to thermal motion of electrons and electrons'
recoil. The former is a symmetric Doppler broadening with
$\mathrm{FWHM}= \sqrt{8\mathrm{ln}2}(kT_{\rm e}/m_{\rm
  e}c^2)^{1/2}\epsilon_0\simeq (0.2\ \mathrm{keV})T_{\rm e, 6}^{1/2}$,
where we have used the rest line energy $\epsilon_0\simeq 7\rm\ keV$.  
This means that the narrow line emitting region must have temperature
$T_{\rm e}\leq 10^6\rm\ K$. The latter (electrons' recoil) causes K$\alpha$
photons to consistently lose energy; $\epsilon_{\rm 0}^2/(m_{\rm
  e}c^2) \simeq 0.1\rm\ keV$ per scattering (assuming $4kT_{\rm e}\ll
\epsilon_0$). This means that the K$\alpha$ photons in this narrow
line are produced in a region where the Thomson depth is $\tau\lesssim
2$.

\subsection{Bouncing between the funnel wall surface}
While the narrow line in the flux spectrum of {\it XMM-Newton}
has FWHM $\sim$0.2 keV, the line in the lag-energy spectrum is
much broader, with FWHM
$\sim$1 keV\footnote{A Gaussian fit with standard deviation 0.67 keV
  and FWHM = 1.6 keV was obtained by \citet{2016Natur.535..388K}, but
  the fit was poor due to the asymmetric line shape (strong red wing
  with little blue wing). The point with half of the peak lag time is
  about 1 keV below the peak energy.}. We show that the broad red wing
in the lag-energy spectrum is likely due to K$\alpha$ photons bouncing
between the inner surface of the envelope.

In \S 4.2, we only considered thermal and Compton broadening
within the K$\alpha$ emitting region. The
average fractional energy loss of a photon in each Compton scattering
is $\epsilon_{\rm 0}/(m_{\rm e}c^2) \simeq 1.4\%$. 
In the physical situation, K$\alpha$ photons are produced at a range
of optical depths $\tau\in$ (1, $\sim$3), so we expect a narrow
component from directly escaping photons which are only Doppler
broadened due to different LOS velocities and a broad component
which is broadened due to both Doppler effect and Compton
scattering \citep[see Fig. 14 of][]{1983ASPRv...2..189P}.
A fraction of K$\alpha$ photons may scatter off electrons up to
$\sim5$-$10$ times (depending on $\tau$ at the emitting location),
possibly causing broadening at the level of $\sim 10\%$ (or 1 keV) toward the red
wing. Due to the small mean free path $(\kappa_{\rm T}\rho)^{-1}\simeq
1.5\times10^{10}(\rho/\bar{\rho})^{-1}\rm\ cm$ ($\bar{\rho}$ is the
mean density of the envelope given by eq. \ref{eq:9}), scattering
within the K$\alpha$ emitting region does not introduce significant
extra lag time. As we will show in section \ref{sec:lag}, the observed
lag time in each energy bin is the proportional to the true lag time
multiplied by the line flux. Thus, if different energy bins have
nearly the same true lag time, the line profile in the lag spectrum is
similar to that in the flux spectrum. In this sub-section, we show that
K$\alpha$ photons bouncing between the inner surface of the funnel
wall also cause broadening but with significant extra lag.

Consider a conical wall of opening angle $\alpha_{\rm c}\lesssim 30^{\rm o}$
moving radially outwards at velocity $\beta$ (and $\gamma =
1/\sqrt{1-\beta^2}$). At point $\mathcal{A}$, we assume that a
K$\alpha$ photon is emitted along the local 
surface normal. The energy of this photon in the BH 
frame is $\epsilon = \epsilon_0/\gamma$, where $\epsilon_0$ is the
rest line energy. We assume that the photon is immediately reflected
when it reaches 
the other side of the wall at point $\mathcal{B}$. The
angle between the momentum vectors of the photon and the electron at
point $\mathcal{B}$ is $\phi = \pi/2 - 2\alpha_{\rm c}$. In the
electron's comoving frame, the photon's energy is $\epsilon' =
\epsilon\gamma (1-\beta \cos\phi)$. We assume that the photon is scattered
toward the direction of the observer (along the jet axis) and ignore
electron's recoil, so the scattered photon's energy in the BH rest
frame will be 
\begin{equation}
  \label{eq:23}
  \epsilon_{\rm obs}^{\mathcal{B}} =
{\epsilon'\over \gamma(1-\beta\cos\alpha_{\rm c})} =
{\epsilon_0(1-\beta\sin 2\alpha_{\rm c}) \over \gamma
  (1-\beta\cos\alpha_{\rm c})}.
\end{equation}
Note that if the photon were initially emitted from point $\mathcal{A}$
directly toward the observer, its energy is
\begin{equation}
  \label{eq:5}
  \epsilon_{\rm obs}^{\mathcal{A}} = {\epsilon_0\over
    \gamma(1-\beta\cos\alpha_{\rm c})}.
\end{equation}
From eqs. (\ref{eq:23}) and (\ref{eq:5}), we see that the energy of
the photon after bouncing ($\epsilon_{\rm obs}^{\mathcal{B}}$) is a
factor of $(1-\beta\sin 2\alpha_{\rm c})$ smaller than that without
bouncing ($\epsilon_{\rm obs}^{\mathcal{A}}$). For typical parameters 
($\alpha_{\rm c}\sim20^{\rm o}$-$30^{\rm o}$, $\beta\sim0.1$-0.2), we
have $\epsilon_{\rm obs}^{\mathcal{A}}/\epsilon_{\rm
  obs}^{\mathcal{B}}-1\in (6\%,\ 17\%)$. We assume point $\mathcal{A}$ is at
distance $R$ from the BH, so the 
extra lag due to bouncing in this case is
\begin{equation}
  \label{eq:7}
  \Delta t_{\rm lag} = {R\over c} \mathrm{tan}2\alpha_{\rm c}(1-\sin\alpha_{\rm c}).
\end{equation}
For $\alpha_{\rm c} = 20^{\rm o}$ ($30^{\rm o}$), we obtain $\Delta
t_{\rm lag}\simeq 1.8\times10^3R_{14}$~s ($2.9\times10^3R_{14}$~s).

The estimates above are based on the assumption that the photon
is emitted along the surface normal at point $\mathcal{A}$. To model
the line profile in detail, one needs to consider a more realistic
geometry with an extended emitting region and then take into account
beaming at the emitting point and electrons' recoil (and possible
absorption) at the scattering point. Instead of going into 
these details, here we only comment on the
qualitative picture that each bounce broadens the line by $\sim 1$~keV
and causes extra lag time of $\sim2000$ seconds. Later in section
\ref{sec:lag}, we show that the red wing of the lag-energy spectrum
from {\it XMM-Newton} is qualitatively consistent with broadening due
to bouncing.

\subsection{Blueshift due to the radiation-driven wind}
Absorption and scattering of $\gamma$/X-rays striking the surface of the
envelope also deposit into the surrounding gas a component of momentum
parallel to the surface. This momentum accelerates a layer of gas near
the surface of the envelope with Thomson depth $\tau_{\rm atm}\sim 3$
(what we call the {\it atmosphere}, see Fig. \ref{fig:transverse}), and
the rate of acceleration is
\begin{equation}
\label{eq:accel}
g_{\parallel} = {\eta_{\rm rel} L\sin^2\theta \over 4\pi \varrho^2}
{\kappa_T \over c\tau_{\rm atm}
  } \cos (\theta - \alpha_{\rm c}), 
\end{equation}
where $L$ is the isotropic-equivalent bolometric luminosity (including
$\gamma$-rays), $\kappa_T$ is the Thomson opacity and 
$\varrho$ is the  cylindrical radius of the 
gas from the jet axis. We assume the acceleration operates from $R =
\varrho/\sin\alpha_{\rm c}$ to $2R$ and obtain the 
asymptotic speed $v \simeq \left(g_{\parallel}R\right)^{1/2}$, i.e.
\begin{equation}
\label{eq:asymp_speed}
v \simeq 0.3c
\left(\frac{\eta_{\rm
      rel}}{0.1}\frac{L_{47.5}}{R_{14}}\frac{3}{\tau_{\rm
      atm}}\right)^{1/2} \frac{\sin\theta}{\sin\alpha_{\rm c}}
  \left[\cos (\theta - \alpha_{\rm c})\right]^{1/2}. 
\end{equation}
Since $\alpha_{\rm c} \simeq \theta\ll1$, the trigonometry factor is
$\simeq1$. Therefore, we find that the $\gamma$/X-ray flux required to
generate the K$\alpha$ photons can naturally accelerate the
K$\alpha$-emitting gas to the speed required by the K$\alpha$
blueshift, $0.1$-$0.2c$. The gas at 
the funnel wall may be further accelerated to higher speeds after
passing through the K$\alpha$ production region.

\section{K$\alpha$ lag time}\label{sec:lag}
The lag-energy spectrum of Swift J1644+57 shows that the variations of
Fe K$\alpha$ line flux at frequencies $\nu_{\rm
  var}=(2$-$10)(1+z)\times10^{-4}\rm\ Hz$ follow the corresponding
variations of the continuum at 4-5 keV 
and 8-13 keV by a time lag of $t_{\rm obs}\sim 120/(1+z)\rm\
s$ (with 1-$\sigma$ error $\sim$40 s). The requirement that the fluorescing gas must be at a distance
$\gtrsim 10^{14}$~cm seems in conflict with this short lag time. 
However, in this section, we show that the measured lag time suffers
from significant dilution due to the coherently varying continuum
component. The dilution-corrected lag time constrains the radius where
the jet radiation is produced.

We denote
the ratio between the K$\alpha$ flux and the direct continuum flux in
the peak lag-time energy bin 7.4-8.1 keV as $q$, and the dilution of
lag time depends on $q$. Since the direct
continuum flux dominates each energy bin, we have $q\ll 1$. Consider
the superposition of a continuum component $\sin\omega t$ and 
a reflected component delayed by $t_{\rm true}$ with a much smaller
amplitude $q\sin\omega (t - t_{\rm true})$, i.e.
\begin{equation}
  \label{eq:75}
  \sin\omega t + q\sin\omega (t - t_{\rm true}) = A\sin\omega(t -
  B t_{\rm true}).
\end{equation}
It is straightforward to show that
\begin{equation}
  \label{eq:12}
  \begin{split}
    &A = (1 + 2q\cos \omega t_{\rm true} + q^2)^{1/2},\\
    &B = \frac{1}{\omega t_{\rm true}}\tan^{-1}\frac{q\sin \omega
      t_{\rm true}}{1 + q\cos \omega t_{\rm true}},
  \end{split}
\end{equation}
and $B$ is the ratio between the observed (diluted) lag time $t_{\rm
  obs}$ and the true lag time $t_{\rm true}$. Note that a
component that does not vary in the (2-$10)(1+z)\times10^{-4}\rm\ Hz$
frequency range has no effect on the measured lag, because it is
filtered out in the cross-correlation process in this specific
frequency range \citep{2014A&ARv..22...72U}. In the limit of
$q\ll 1$, we have
\begin{equation}
  \label{eq:36}
  \frac{t_{\rm obs}}{t_{\rm true}} = B \simeq q
  \mathrm{sinc}(2\pi \nu_{\rm var} t_{\rm true}) 
\end{equation}
where $\nu_{\rm var}$ is the frequency of flux variations and we have
used the function $\mathrm{sinc}(x) = \sin x/x$.

In section \ref{sec:Kalpha}, we have estimated $EW\sim
400\bar{\eta}_{\rm rel}\rm\ eV$ with an uncertainty of a factor of
$\lesssim 3$, where $\bar{\eta}_{\rm rel}$ is the X-ray
intensity correction due to relativistic beaming averaged
over the reprocessing surface and $\bar{\eta}_{\rm rel}\lesssim 
0.2$ is required by the Fe ionization state. Hereafter, we suppose that
the EW of the broad K$\alpha$ line is $\lesssim 80\rm\ eV$, so the
ratio between the line and continuum fluxes in the 7.4-8.1 keV energy
bin has an upper limit (from assuming all K$\alpha$ photons to be in
this bin)
\begin{equation}
  \label{eq:73}
  q \lesssim \frac{EW}{0.7\rm\ keV} \lesssim 0.1
\end{equation}
Since $\mathrm{sinc}(2\pi \nu_{\rm var} t_{\rm true}) < 1$, we obtain
from eq. (\ref{eq:36}) a lower limit of the true lag time $t_{\rm true}
\gtrsim 8\times10^2\rm\ s$.

In the lag-energy spectrum, the measured lag time drops toward the
red wing. Broadening due to bouncing between the funnel wall surface
causes longer lag time, and 
hence the flux ratio $q$ must decrease toward the red 
wing faster than the increasing of lag time. From \S 4.3, we see that
bouncing between the funnel wall surface causes broadening of $\sim
1$~keV and extra lag time $\Delta t_{\rm
  lag}\sim2\times10^{3}$~s. The lag-energy spectrum has rather poor
energy resolution and the half-lag point (where the measured lag is
half of the peak lag) is roughly $\sim1$~keV below
the peak. The line-to-continuum flux ratio at the half-lag point
$q_{1/2}$ is smaller than the ratio at the peak $q$
(eq. \ref{eq:73}) by a factor of $t_{\rm  
true}(\mathrm{peak})/2\Delta t_{\rm lag}\gtrsim 0.2$ (using $t_{\rm
true}(\mathrm{peak}) \gtrsim 
8\times10^2\rm\ s$), i.e. $q_{1/2}\gtrsim 0.2q$. This is qualitatively
consistent with the fact 
that broadening due to bouncing decreases the line flux
density. 

Modeling the lag-energy spectrum in detail is out of the
scope of current work, but we emphasize the following two points: (1)
scattering off the funnel wall causes photon energy redshift from the
line center and introduces extra lag time; (2) the line amplitude in
the lag-energy spectrum is proportional to the product of the number of
K$\alpha$ photons and their lag time, while the flux-spectrum
amplitude is proportional only to the number of K$\alpha$
photons. Therefore, the line in the lag-energy spectrum is broader
than that in the flux spectrum. 

\begin{figure}
  \centering
\includegraphics[width = 0.3 \textwidth,
  height=0.33\textheight]{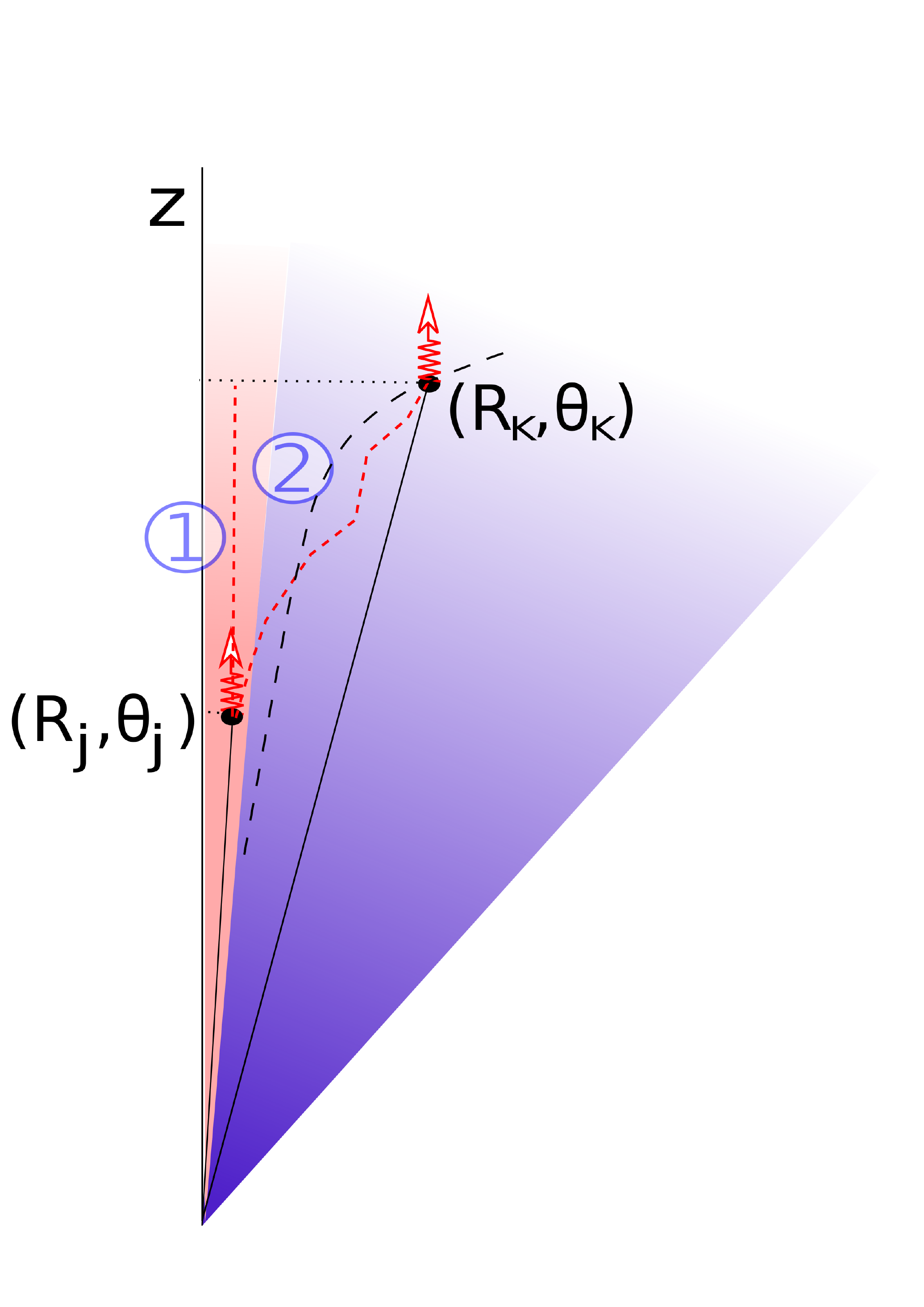}
\caption{Schematic picture of Fe K$\alpha$ line production in
  the jet-envelope interaction geometry. We use spherical polar
  coordinates centered on the BH with polar/z axis pointing
  toward the observer. The positions of the jet radiation and
  K$\alpha$ emission 
  are $(R_{\rm j}, \theta_{\rm j})$ and $(R_{\rm K}, \theta_{\rm K})$,
  respectively. The true lag time between the continuum
  and K$\alpha$ line is light-travel time difference between the two
  paths marked in dashed red curves. The black dashed curve is the
  Thomson surface where the optical depth for electron scattering
  $\tau=1$.
}\label{fig:lag_time}
\end{figure}

In the following, we use the true lag time to
constrain the jet radiation radius. Let us assume the positions of
direct continuum and K$\alpha$ emission 
are $(R_{\rm j}, \theta_{\rm j})$ and $(R_{\rm K}, \theta_{\rm K})$
respectively in spherical polar coordinates as shown in
Fig. (\ref{fig:lag_time}). The light-travel 
time for direct continuum is
\begin{equation}
  \label{eq:14}
t_{1} = (R_{\rm K} \cos \theta_{\rm K} - R_{\rm j}\cos
\theta_{\rm j})/c \simeq (R_{\rm K} \cos \theta_{\rm K} - R_{\rm j})/c
\end{equation}
where we have used $\cos \theta_{\rm j}\approx 1$.
The light-travel time for diffusion plus advection
in the envelope $t_{2}$ depends on the detailed velocity and density
profile of the envelope, but a simple estimate can be made as
follows. In the optically thin limit $\tau\sim
1$, photons travel almost in a straight
line, i.e.
\begin{equation}
  \label{eq:15}
  t_{2} \simeq (R_{\rm K}^2 + R_{\rm j}^2 - 2 R_{\rm K}R_{\rm j} \cos
  \theta_{\rm K})^{1/2}/c, 
\end{equation}
where we have used $\theta_{\rm j}\ll \theta_{\rm K}$. Using
$R_{\rm K} = xR_{\rm j}$, we have 
\begin{equation}
  \label{eq:21}
  t_{2} - t_{1} \simeq (\sqrt{x^2 - 2x\cos\theta_{\rm K} + 1} - x\cos
  \theta_{\rm K} + 1)R_{\rm j}/c,
\end{equation}
which depends on $x$ weakly. It varies in $(0.25, 0.5)R_{\rm
  j}/c$ or $(0.5, 1.0)R_{\rm j}/c$ when $x\in (1.5, 7)$ for
$\theta_{\rm K} = 20^{\rm o}$ or $30^{\rm o}$ respectively, so we
obtain 
\begin{equation}
  \label{eq:29}
   t_{2} - t_{1} \simeq R_{\rm j}/2c,\ \mathrm{if}\ \tau
   \sim 1.
\end{equation}
On the other hand, when the optical depth of the line production
region is large $\tau \gg 1$, photons undergo many scatterings
$N\sim\tau$ and hence travel slower than free streaming. Photons are 
either advected by the fluid motion (if $\tau > 1/\beta$) 
or diffuse through the envelope (if $\tau < 1/\beta$), and in any
of the two cases, a conservative estimate of the travel time is the diffusion
time\footnote{Consider a 
  photon in a 3D random walk with step sizes in Poisson distribution of
  mean value $\lambda$ (mean free path). The mean distance traveled
  after N steps is $S_{\rm N} = \sqrt{2N}\lambda$ and the mean travel
  time is $t_{\rm N} = N\lambda/c = S_{\rm N}^2/(2\lambda c)$. }
\begin{equation}
  \label{eq:19}
    t_{2} \simeq \tau R_{\rm j}/2 c,\ \mathrm{if}\ \tau \gg 1.
\end{equation}
Combining eq. (\ref{eq:29}) and (\ref{eq:19}), we obtain the true
lag time expected from the jet-envelope interaction
\begin{equation}
  \label{eq:37}
  t_{\rm true} = t_{2} - t_{1} \simeq \tau R_{\rm j}/2c,\
  \mathrm{for\ any\ }\ \tau \gtrsim 1,
\end{equation}
which gives $\tau R_{\rm j}\simeq 5\times 10^{13}(t_{\rm
  true}/8\times10^2\rm\ s)\rm\ cm$. Since $t_{\rm true}\gtrsim
8\times10^2\rm\ s$ and $\tau\sim3$, we obtain a lower limit of the
jet radiation radius of $\sim2\times 10^{13} 
\rm\ cm$. On the other hand, as we show in Appendix A, the
photospheric radius of the envelope is typically $\sim$ a
few$\times10^{14}\rm\ cm$, which 
gives an upper limit of the jet radiation radius of
$\sim 3\times10^{14}\rm\ cm$. We conclude 
that the radius where the jet energy is converted to radiation is
$2\times 10^{13}\lesssim R_{\rm j}\lesssim 3\times 10^{14}\rm\ cm$.

\section{Discussion}\label{sec:discussion} 
In this section, we discuss the implications of our results for a
  variety of topics, including
the radiation mechanism of $\gamma$/X-rays, other observable
signatures of jet-envelope interaction, and a possible explanation of
the late-time radio emission from Swift J1644+57. Some remaining
issues and caveats are mentioned at the end.

(1) Many different radiation mechanisms for the $\gamma$/X-ray
continuum in 
Swift J1644+57 have been discussed by
\citet{2016MNRAS.460..396C}. The 
leading candidates are synchrotron emission from  
electrons accelerated by magnetic reconnection and the external
radiation being inverse-Compton scattered by electrons in the jet
(hereafter EIC model). \citet{2016MNRAS.458.1071L} showed that the EIC
model is consistent with Swift J2058+05 where the thermal external
radiation field was observed in the UV-optical. An issue of the EIC
model is how to produce a power-law $\gamma$/X-ray spectrum, because
hot electrons cool down to Lorentz factors $\gamma_{\rm e}\sim 1$
extremely quickly. Here we discuss the implications of the results from
this work on the EIC model. The general idea is that multiple
scattering within a flow that has strong shear motion can produce a
power-law spectrum in a Fermi-like process. The number of leptons per
proton\footnote{We assume that the jet kinetic 
energy is dominated by protons. If the kinetic energy is
dominated by $e^{\pm}$ pairs, the optical depth for annihilation at
the base of the jet is $\sim \chi m_{\rm p}/(\sigma_0 m_{\rm 
  e})$, where $\sigma_0$ is the initial magnetization and $\chi$ is
the ratio between true jet power and the 
Eddington luminosity and we have assumed Thomson cross-section
because relativistic leptons cool rapidly via synchrotron process. For
Swift J1644+57, we have $\chi \sim10$-$100$ (depending on the
radiation efficiency), so pairs will annihilate after jet launching if
$\sigma_0 \lesssim 10^4$. Pair plasma also suffers from a much
stronger Compton drag that may slow down the jet quickly.} is denoted
as $\xi_{\rm e}\in[1, m_{\rm p}/m_{\rm 
  e}]$ and the magnetization parameter is $\sigma$. If the jet has
isotropic equivalent power $L_{\rm j} = 10^{48} L_{\rm j, 48}\rm\ erg\
s^{-1}$, Lorentz factor $\Gamma = 10\Gamma_1$ and
half-opening angle $\theta_{\rm j} = 0.1 \theta_{\rm j, -1}$,
electron number density at radius $R = 10^{14}R_{14}$~cm is $n_{\rm e}
= L\xi_{\rm e}/[4\pi R^2 \Gamma m_{\rm p}c^3 \mathrm{max}(1,\sigma)]$
(in the BH's rest frame), so the jet
optical depth in the transverse direction is 
\begin{equation}
  \label{eq:8}
  \tau_{\rm j,trvs} \simeq \sigma_{\rm T}n_{\rm e} R\theta_{\rm j}
  \simeq 0.12\frac{L_{\rm j,48}
    \theta_{\rm j,-1} \xi_{\rm
      e}}{R_{14}\Gamma_1\mathrm{max}(1,\sigma)}, 
\end{equation}
and the optical depth for a photon propagating in the radial
direction outward is a factor of $\sim 10$ smaller.

The fraction of the jet radiation impinging the
surface of the envelope is
\begin{equation}
  \label{eq:13}
  f_{\rm imp} = \frac{\int_{\alpha_{\rm c}}^{\pi/2} \frac{\d
      L}{\d \Omega}(\theta)
    \sin\theta \d \theta }{\int_0^{\pi/2} \frac{\d
      L}{\d \Omega}(\theta)
    \sin\theta \d \theta}\sim \bar{\eta}_{\rm rel}.
\end{equation}
This estimate is a consequence of the fact that $\d
L/\d\Omega (\theta)$ is a rapidly-declining function of 
$\theta$. Because $\eta_{\rm rel}(\theta) \equiv [\d L/\d\Omega
(\theta)]/(\d L/\d\Omega|_{\rm LOS})$ is the X-ray intensity
correction due to 
relativistic beaming, we have $\int_{\alpha_{\rm c}}^{\pi/2}\d
L/\d \Omega \sin\theta \d \theta \sim \d L/\d\Omega|_{\rm LOS}\
\bar{\eta}_{\rm rel} \alpha_{\rm c}^2/2$. The LOS of the observer is close to
the jet axis and most radiation power is beamed within the opening
cone of angle $\alpha_{\rm c}$, so the denominator is $\sim
\d L/\d\Omega|_{\rm LOS}\ \alpha^2_{\rm c}/2$, and hence $f_{\rm
  imp}\sim \bar{\eta}_{\rm rel}$. This means roughly a fraction of
$\sim 10\%\bar{\eta}_{\rm rel, -1}$ of the beaming corrected jet
radiation energy is reprocessed by the surrounding envelope. 

Consider a fraction $f_{\rm imp}\sim \bar{\eta}_{\rm rel}$ of the
photons that have been scattered once by the jet impinging on the
surrounding gas again. When 
the ionization parameter is 
high $\xi>10^3\rm\ erg\ cm\ s^{-1}$, the albedo in the
0.3-10 keV range of is 
of order unity \citep{1988ApJ...335...57L}, so most of the photons are
reflected back to the jet region and will likely get scattered again
by the jet electrons. If electrons in the jet are cold
($\gamma_{\rm e}\sim 1$), photons scattered by the jet obtain a
fractional energy gain of $\Gamma_{\rm r}^2$, where $\Gamma_{\rm
  r}\simeq \Gamma \Gamma_{\rm s} (1 - \beta_{\rm s})$ is the
relative Lorentz factor 
between the jet (Lorentz factor $\Gamma$) and the reflecting gas at
the Thomson surface (speed $\beta_{\rm s}$ and Lorentz factor
$\Gamma_{\rm s}$). 

Therefore, the Compton $y$ parameter for repetitive scattering between
the jet and surrounding envelope is
\begin{equation}
  \label{eq:74}
  y \simeq \bar{\eta}_{\rm rel} \tau_{\rm j,trvs}\Gamma_{\rm r}^2 \simeq 0.6
  \frac{\bar{\eta}_{\rm rel, -1} L_{\rm j,48}\xi_{\rm
      e}}{R_{14}\mathrm{max}(1,\sigma)} \frac{\Gamma_{\rm s}^2(1 -
    \beta_{\rm s})^2}{0.5} (\Gamma\theta_{\rm j}), 
\end{equation}
from which we see that $y$ is of order unity for a baryonic jet at radius 
$R\sim 10^{14}\rm\ cm$, so a power-law spectrum may be
produced in the EIC model even when the jet electrons are cold. Recall
that in the model of \citet{2016Natur.535..388K} the source is moving
sub-relativistically, and relativistic power-law electrons are
required to produce the observed hard power-law $\gamma$/X-ray 
spectrum. The thermal seed photons from the surrounding envelope have
energy density $\sim aT^4\sim 10^6 \rm\ erg\ cm^{-3}$ for $T =
10^5\rm\ K$, which means the inverse-Compton cooling time of an
electron with Lorentz factor $\gamma_{\rm e}\gg 1$ is very short:
$\sim 30\gamma_{\rm e}^{-1}\rm\ s$. Here, 
the EIC model with {\it cold} electrons does not have the cooling
problem. The isotropic equivalent EIC luminosity from a baryonic jet
with cold electrons is given by
\begin{equation}
  \label{eq:24}
  \begin{split}
      L_{\rm EIC} &\sim {L_{\rm j} R/c \over \Gamma m_{\rm p} c^2} \times
  \sigma_{\rm T} c \Gamma^2 aT^4 \\
  &\sim (3\times10^{47} \mathrm{\ erg\ s^{-1}})\ L_{\rm j, 48} R_{14}
  \Gamma_1 T_5^4,
  \end{split}
\end{equation}
which is consistent with the observed $\gamma/$X-ray
luminosity in the first $\sim 10\rm\ d$ \citep{
2011Natur.476..421B, 2011Sci...333..203B}. Note that the jet optical
depth in the transverse direction ($\tau_{\rm j, trvs}$,
eq. {\ref{eq:8}}) drops dramatically when
photon energies in the jet comoving frame approach the Klein-Nishina
regime ($h\nu'\sim m_{\rm 
  e}c^2$), so the power-law spectrum cuts off at $\sim$a few MeV
in the EIC model. This could be tested by future observations.

(2) In order for the surface of the envelope not to be over-ionized, the ratio
between the X-ray intensity striking the surface and the intensity
beamed in the direction of the observer is
$\bar{\eta}_{\rm rel}\lesssim 0.2$. However, we did not specify how
the angle-dependent radiation power distribution $\d L/\d
\Omega(\theta)$ is 
physically realized. One scenario is that 
the jet has a core-sheath structure, i.e. an ultra-relativistic core is
surrounded by a  mildly relativistic sheath with $\Gamma_{\rm sh}\sim
2$ \citep[e.g.][]{2003ApJ...594L..27G, 2004ApJ...600..127G,
  2005A&A...432..401G}, so the sheath scatters the narrowly beamed
radiation from the core to much wider angles. 
Another possibility is that the jet has Lorentz factor of $\Gamma\sim
3$-5 but the opening angle of 
the jet is much narrower than $1/\Gamma$. This allows a significant
fraction of jet radiation to impinge on the surrounding optically
thick gas, and at the same time direct hydrodynamical contact between
the jet and envelope (baryon pollution) is avoided. The third scenario
is that, although the bulk motion of the jet has a 
large Lorentz factor $\Gamma \sim 10$, magnetic reconnection at the
jet dissipation radius produces relativistic motion in the jet
comoving frame \citep[e.g.][]{2009MNRAS.394L.117N,
  2009MNRAS.395L..29G}. This 
allows the relativistic beaming angle to be much wider 
than $1/\Gamma$. The results in this work do not depend on how the
angle dependence of the jet radiation is physically realized, as long
as the beaming correction $\bar{\eta}_{\rm rel}\lesssim 0.2$.

(3) The total amount of radiation power reprocessed by
the surrounding envelope is given by
\begin{equation}
  \label{eq:10}
  \begin{split}
      L_{\rm imp} & = f_{\rm imp} \int_0^{\pi} \frac{\d
        L}{\d \Omega}(\theta) \sin \theta \d \theta \sim
      \bar{\eta}_{\rm rel} \alpha_{\rm c}^2 \left. \frac{\d L}{\d
        \Omega}\right|_{\rm LOS}\\
 & \sim (7\times 10^{44}\rm\ erg\ s^{-1})
   \left(\frac{\alpha_{\rm c}}{30^{\rm o}}\right)^2 \bar{\eta}_{\rm rel,
     -1} L_{47.5},
  \end{split}
\end{equation}
where we have used
$ \int_0^{\pi} \d L/\d \Omega \sin \theta \d \theta\sim 
\d L/\d\Omega|_{\rm LOS}\ \alpha^2_{\rm c}$, $f_{\rm imp}\sim
\bar{\eta}_{\rm rel}$, and
$L = 4\pi\ \d L/\d 
\Omega|_{\rm LOS} = 10^{47.5}L_{47.5}\rm\ erg\ 
s^{-1}$ is the isotropic equivalent bolometric luminosity from the
jet. In the 
momentum-driving 
limit (eq. \ref{eq:accel}), the expanding atmosphere only carries a small
fraction of the energy of the reprocessed jet radiation. The
$\gamma$/X-rays may be advected out with the outflow or bounce between
the inner walls until the photons' momenta are directed within the
narrow opening angle of the funnel wall, and in this
process they could get absorbed (bound-free absorption) or Compton
down-scattered. For example, a photon
of energy $\epsilon = m_{\rm e}c^2/N$ loses most of its energy to
electrons after $\sim N$ scatterings. The energy of the 
reprocessed jet radiation is converted to gas internal energy which 
is then converted to bulk kinetic energy via adiabatic expansion. This
energy-driven wind likely involves more envelope mass and is wider
than the momentum-driven one. The power of the energy-driven wind
could be a large fraction of $L_{\rm imp}$ given in eq. (\ref{eq:10}),
and hence it may be much stronger than the possible wind
launched from the accretion disk. 
If all jetted TDEs have jet-envelope interactions similar to the one
we propose for Swift
J1644+57, their UV-optical emission may be systematically brighter 
than in non-jetted TDEs. If the observer's LOS is
within the cone of the reprocessing wall, the thermal UV-optical
emission due to the reprocessed jet radiation has temperature $T\sim 
[L_{\rm imp}/(4\pi R^2\sin^2 \alpha_{\rm c}\sigma_{\rm SB})]^{1/4}
\sim (1.4\times 10^5\rm K)\ \bar{\eta}_{\rm rel, -1}^{1/4} L_{47.5}^{1/4}
R_{14}^{-1/2}$, where $\sigma_{\rm SB}$ is
the Stefan-Boltzmann constant. The temperature and
luminosity will be lower for off-axis observers.

(4) The radio emission from Swift J1644+57 was from the external shocks
produced when the jet and possibly a wind interacted with the
circum-nuclear medium (CNM) at large distances from the BH. 
The radio lightcurve and spectrum in the first few weeks can be
adequately explained by the external shocks from a relativistic jet of
Lorentz factor $\sim 10$ \citep{2012MNRAS.420.3528M}. In their model,
the CNM 
density decreases with radius rapidly $n\propto r^{-2}$. 
Since no indication of significant late-time energy injection into the
jet was seen in the X-ray lightcurve, as the jet-driven shock
decelerates, its radio emission should fade away after $\sim 30\rm\
d$. However, an unexpected rebrightening phase
was found at $\sim 150\rm\ d$ post-discovery, which 
may be due to a slower non-/mildly-relativistic wind
\citep{2012ApJ...748...36B, 2013ApJ...767..152Z,
  2013ApJ...770..146B,2015MNRAS.450.2824M}, 
although \citet{2013MNRAS.434.3078K} had a different
explanation. Here, we propose that the
jet-envelope interaction naturally launches a wide-angle
non-relativistic wind, which drives an external forward shock capable
of explaining the radio rebrightening.

The emission from the forward shock
roughly peaks at the deceleration radius $R_{\rm dec}$ where the
swept-up mass equals to the ejecta mass. For the purpose of a rough 
estimate, we assume the CNM density profile in Swift J1644+57 to be
similar to our Galactic Center $n(R) = n_{\rm
  pc}(R/\mathrm{pc})^{-1.5}$ and $n_{\rm pc} = 10 n_{\rm 
  pc,1}\rm\ cm^{-3}$ \citep{2003ApJ...591..891B}. If the wind has
kinetic energy $E_{\rm 
  w}$ and speed $\beta_{\rm w}c$, the deceleration radius is
$R_{\rm dec} \simeq (0.03\rm\ pc)\ E_{\rm w, 51}^{2/3} (\beta_{\rm
  w}/0.3)^{-4/3} n_{\rm pc,1}^{-2/3}$, so the deceleration time is
$t_{\rm dec}\simeq R_{\rm dec}/(\beta_{\rm w}c)\simeq (100\rm\ d)\
E_{\rm w, 51}^{2/3} (\beta_{\rm w}/0.3)^{-7/3} n_{\rm
  pc,1}^{-2/3}$. The emission from external shocks has been well studied
in the gamma-ray burst literature and it is generally assumed that
power-law electrons and magnetic fields share 
fractions of $\epsilon_{\rm e}$ and $\epsilon_{\rm B}$ of the thermal
energy in the shocked region \citep[e.g.][]{2015PhR...561....1K}.
The forward shock accelerates swept-up
electrons to a power-law distribution $\d N/\d \gamma_{\rm e}\propto
\gamma_{\rm e}^{-p}$ with minimum Lorentz factor $\gamma_{\rm m} = 
(p-1)\epsilon_{\rm e} m_{\rm p} \beta_{\rm w}^2/[2(p-2) m_{\rm
  e}]\simeq 30\ \epsilon_{\rm e, -1} (\beta_{\rm w}/0.3)^2$ for $p =
2.4$, which is given by the high-frequency radio spectrum
$F_{\nu}\propto  \nu^{-0.7}=\nu^{(1-p)/2}$. At the
deceleration radius, the magnetic field strength in  
the shocked region is $B = (0.1\rm\ G) \epsilon_{\rm B, -2}^{1/2}
(\beta_{\rm w}/0.3)$. The synchrotron emission from an electron
with Lorentz factor $\gamma_{\rm e}$ peaks at $\nu = 3\gamma_{\rm
  e}^2eB/(4\pi m_e c) \simeq  (4.2\rm\ GHz) \gamma_{\rm e, 2}^2B_{-1}$
and the peak specific power is $P_{\rm \nu, peak}\simeq e^3
B/(m_{\rm e}c^2)$.

The late-time radio spectrum peaks at $\sim 10$~GHz and roughly has
$F_{\nu}\propto \nu^2$ below the peak frequency
\citep{2012ApJ...748...36B}. The synchrotron 
frequency associated with 
$\gamma_{\rm m}$ is much below GHz, so the peak corresponds to the
self-absorption frequency $\nu_{\rm a}$ (associated with Lorentz 
factor $\gamma_{\rm a}$) roughly given by $4\pi R_{\rm dec}^2\
2 kT_{\rm a}\nu_a^2/c^2 \simeq N(\geq 
\gamma_{\rm a}) P_{\rm \nu, peak}/(4\pi)$, where $kT_{\rm a} =
\gamma_{\rm a}m_{\rm e}c^2$ is the temperature associated with
the emitting electrons. The number of electrons emitting at
frequency $\nu_{\rm a}$ is $N(\geq \gamma_{\rm a}) \simeq
2E_{\rm w}/(\beta_{\rm w}^2m_{\rm p} c^2)\ (\gamma_{\rm a}/\gamma_{\rm
  m})^{1-p}$. Therefore, we obtain the self-absorption frequency
$\nu_{\rm a}\simeq (6\rm\ GHz)\ E_{\rm w, 51}^{-0.1} (\beta_{\rm
  w}/0.3)^{1.8} n_{\rm pc,1}^{0.4} \epsilon_{\rm B,
  -2}^{0.7}\epsilon_{\rm e, -1}^{0.4}$ and the peak flux density
$F_{\nu_{\rm a}}\simeq (7\rm\ mJy) \ E_{\rm w, 51}^{1.1} (\beta_{\rm 
  w}/0.3)^{1.3} n_{\rm pc,1}^{-0.3} \epsilon_{\rm B, 
  -2}^{0.6}\epsilon_{\rm e, -1}^{1.1}$ (for z = 0.354). To within a
factor of $\lesssim 3$, these estimates roughly agree with the radio data 
after about $90\rm\ d$  \citep{2012ApJ...748...36B} when the
jet-driven  external shocks have faded away. Better agreement can be
comfortably made by 
adjusting some of the parameters such as $\epsilon_{
\rm B}$ and $\epsilon_{\rm e}$. For more detailed modeling, one needs
to consider many uncertain factors: the angular 
distribution of wind energy/speed, 
the unknown CNM density profile, and electron acceleration and
magnetic field amplification at collisionless shocks.

(5) Due to the uncertainty and complexity of the system, our
jet-envelope interaction model is
limited in the following two aspects. First, we only consider the impact of
the jet radiation on the surrounding envelope and have ignored the
feedback on the jet. In reality, the jet and envelope may be
dynamically coupled through the effective viscosity provided by 
Compton scattering. Second, the detailed K$\alpha$ line
production and radiative transfer processes (e.g. line flux profile,
lag-energy profile) have not 
been studied. Our model can 
be improved by future numerical simulations of jet-envelope
coupling and will be tested by future observations of jetted TDEs.

\section{Conclusion}\label{sec:summary} 

If the $\gamma$/X-rays from jetted TDEs
are produced below the photospheric radius of the surrounding
optically thick envelope, a fraction of the jet radiation may interact
with and affect the dynamics of the envelope. An evidence
of this interaction is the Fe K$\alpha$ line detected in the
reverberation lag spectrum from Swift J1644+57 by
\citet{2016Natur.535..388K}.

The discovery paper argued that the
source of the $\gamma$/X-ray continuum was located very close to the
BH ($\sim30$ gravitational radii) and moved sub-relativistically. We
have reanalyzed the lag spectrum, pointing out that dilution effects cause 
it to indicate a geometric scale an order of magnitude larger than
inferred by \citet{2016Natur.535..388K}. If the $\gamma$/X-ray
continuum is produced by a relativistic jet, as 
suggested by the rapid variability, high luminosity and hard spectrum,
this larger scale predicts an Fe ionization state consistent with
efficient K$\alpha$ production. Moreover, the relativistically beamed
jet radiation impinging on the funnel wall of the surrounding gas also
accelerates the reprocessing layer of Thomson depth $\tau\sim 3$ to
speeds $\beta\sim0.1$--$0.2$.

Our model can explain the following
observational results qualitatively: (i) the line energy is blueshifted
from the rest energy $\epsilon_0$ to $\epsilon_{\rm obs} =
\epsilon_0/[\gamma (1-\beta\cos 
\alpha_{\rm c})] \simeq (1+\beta)\epsilon_0$ for
funnel wall opening angle $\alpha_{\rm c}\lesssim30^{\rm o}$ and
outflowing speed $\beta \lesssim0.2$; (ii) 
the broad asymmetric red wing in the lag-energy spectrum extending from
the peak at $\simeq$8 keV down to 6--7 keV (energies in the
host-galaxy rest frame) is due to K$\alpha$ photons bouncing between the inner
surface of the envelope plus Compton down-scattering;
(iii) the K$\alpha$ lag time of
$\sim 120/(1+z)\rm\ s$, which suffers from dilution by the 
coherently varying continuum photons that dominates in each energy bin,
indicates that the jet radiation radius is at $R_{\rm j}\gtrsim2\times
10^{13}\rm\ cm$ (and the photospheric radius of the reprocessing gas
gives an upper limit of $R_{\rm j}\lesssim 3\times10^{14}\rm\ cm$).

Our model fits well with the global picture of jetted TDEs. Although
our model does not depend on the jet radiation mechanism, we show that
the power-law $\gamma$/X-rays may be produced by the external photons
from the surrounding envelope being repetitively inverse-Compton
scattered by {\it 
  cold} electrons in the jet. The reprocessed jet
radiation drives a non-relativistic wind, which may explain
the late-time radio rebrightening of Swift J1644+57. This energy
injection may also cause the thermal UV-optical emission from jetted
TDEs to be systematically brighter than in non-jetted ones.

\section{acknowledgements}
We thank the anonymous referee for many useful comments, which
improved the content and clarity of the paper. 
We thank Erin Kara, Chris Reynolds and Phil Uttley for helpful
discussion on the K$\alpha$ lag spectrum. We thank Christopher Orban
and Lianshui Zhao for help in getting the mono-energetic opacity from
the Opacity Project. The idea of a non-relativistic wind possibly
causing the radio rebrightening was from discussion with Zhuo Li at Peking
University a few years ago. We also thank Sera Markoff for arranging
this collaboration. 
WL is funded by the Named Continuing
Fellowship at the University of Texas at Austin.
JHK was partially supported by NASA grant NNX14AB43G and by NSF grant
AST-1516299. PC acknowledges financial support from the WARP program
of the Netherlands Organisation for Scientific Research (NWO). 


\appendix
\section{ }

\begin{table*}
 \begin{minipage}{0.95\textwidth}
   \centering
  \caption{The photospheric radius defined by $L_{\rm
      bol} = 4\pi R_{\rm ph}^2 \sigma_{\rm SB}T^4$ can be constrained
    by eqs. (\ref{eq:18}) and (\ref{eq:6}) from one data point
    ($\nu_0$, $L_0\equiv \nu_0L_{\nu_0}$)
    on the R-J tail at a certain post-discovery
    time in the host-galaxy rest frame. We list the constraints for some 
    recent TDEs. References: (1) \citet{2015ApJ...805...68P}, (2)
    \citet{2016ApJ...819...51L}, (3)   
    \citet{2009ApJ...698.1367G}, (4) \citet{2012Natur.485..217G}, (5)
    \citet{2016MNRAS.455.2918H}.}
  \label{tab:Rph}
  \begin{tabular}{@{}ccccccc@{}}
  \hline\hline
TDEs& Refs. & $\nu_0$[Hz] &$L_0$[erg/s] & time[d] & $R_{\rm
min}$[cm]&$R_{\rm max}$[cm]\\ 
\hline
Sw J2058+05&(1) &1.0e15&1.3e44 &11 &2.2e14 &4.8e14\\
\hline
Sw J1644+57\footnote{For Swift
    J1644+57, strong dust extinction causes a large uncertainty (up to
    a factor of $\sim$10 to 
    30) in the r-band flux, due to unknown reddening $E(B-V)$ and
    extinction law parameter $R_{\rm 
    V}\equiv A_{\rm V}/E(B-V)$. The uncertainties at longer wavelengths
    are much less ($\sim50\%$ at H-band), but the 
    near-infrared spectrum before and after extinction correction is
    clearly not R-J like \citep[there may be a
    power-law component due to external shocks, see][their 
    Fig. 6]{2016ApJ...819...51L}. Here, for the 
    purpose of a rough estimate, we apply $R_{\rm V} = 3.1$ and $E(B-V) =
    1.5$~mag to the r-band data and correct for dust extinction
    of 5.5 mag in the host-galaxy rest frame, using the
    reddening law calibrated by 
    \citet{2011ApJ...737..103S}.
    For $R_{\rm V} = 3.1$ and $E(B-V)
    =2$~mag, the photospheric radius is between $R_{\rm
    min}=8.5\times10^{14}$~cm and $R_{\rm max} = 1.6\times10^{15}$~cm. 
}&(2)  
            &7.1e14&$\sim$6.4e43 &17& $\sim$2.6e14&$\sim$6.6e14\\ 
\hline
D23H-1&(3) &7.5e14&2.9e42&20 &3.1e13 &1.3e14 \\
\hline
D3-13 &(4) &8.6e14&9.0e42&$\sim$20& 4.9e13&1.7e14\\
\hline
PS1-10jh & (4) &4.2e14 &5.7e42&30& 1.5e14&5.6e14 \\
\hline
ASSASN-14li& (5)&9.1e14&9.2e42&8& 4.3e13&1.5e14\\
\hline\hline
\end{tabular}
\end{minipage}
\end{table*}

We show that the photospheric radius of the envelope responsible for
the thermal UV-optical emission is $\sim$ a few$\times10^{14}\rm\
cm$. We define the photospheric radius $R_{\rm ph}$ based on the two
observable quantities of a thermal spectrum, bolometric luminosity
$L_{\rm bol}$ and temperature $T$, in the following way
\begin{equation}
  \label{eq:2}
  L_{\rm bol} = 4\pi R_{\rm ph}^2 \sigma_{\rm SB} T^4,
\end{equation}
where $\sigma_{\rm SB}$ is the Stefan-Boltzmann constant. Note that
$R_{\rm ph}$ here does not necessarily equal to the Thomson surface
radius $R_{\rm s}$ where the 
optical depth for electron scattering $\tau$ equals to
1. If the bound-free opacity is comparable to electron scattering opacity,
thermalization happens near the Thomson surface $R_{\rm s}\simeq
R_{\rm ph}$.

The peak of the TDE  thermal component (at $h\nu\simeq
2.8kT$) is always in the far UV and has never been observed. Therefore,
there are large uncertainties in $L_{\rm bol}$ and $T$ and eq. (\ref{eq:2})
cannot be directly used to measure $R_{\rm ph}$. Usually, the
observable part is on the Rayleigh-Jeans (R-J) tail
\begin{equation}
  \label{eq:3}
  \nu L_{\nu} = \frac{15L_{\rm bol}}{\pi^4}
  \left(\frac{h\nu}{kT}\right)^3 = L_0 (\nu/\nu_0)^3,
\end{equation}
where we have normalized the spectrum with a measured data point ($\nu = 
\nu_0$, $\nu L_{\nu} = L_0$) on the R-J tail. We put $L_{\rm bol}$
from eq. (\ref{eq:2}) into eq. (\ref{eq:3}), take $\nu = \nu_0$, and
then obtain the temperature $T$ as a function of photospheric radius
$R_{\rm ph}$
\begin{equation}
  \label{eq:20}
  kT = {L_0 c^2 \over 8\pi^2 \nu_0^3} {1 \over R_{\rm ph}^2}
\end{equation}
Since the observed data point is far below the blackbody peak, we have
\begin{equation}
  \label{eq:4}
  h\nu_0 < kT.
\end{equation}
Combining eqs. (\ref{eq:20}) and (\ref{eq:4}) gives an upper limit of $R_{\rm
  ph}$ which we denote as $R_0$, i.e.
\begin{equation}
  \label{eq:18}
  R_{\rm ph} < R_0\equiv \frac{\sqrt{2}L_0^{1/2}c}{4\pi h^{1/2} \nu_0^2}.
\end{equation}
On the other hand, eq. (\ref{eq:3}) also gives us the temperature as a
function of the bolometric luminosity ($\nu_0$ and $L_0$ are knowns)
\begin{equation}
  \label{eq:22}
  kT = {15 h\nu_0 \over \pi^4 L_0} L_{\rm bol}\leq {15 h\nu_0 \over
    \pi^4 L_0} L_{\rm max}, 
\end{equation}
where the maximum bolometric luminosity $L_{\rm max} \sim 10^{47}\rm\
erg\ s^{-1}$ comes from assuming a
total energy budget of $\sim 10^{53}\rm\ erg$ and
the duration of peak luminosity $\sim 10^6\rm\ s$. Then, this upper
bound on temperature gives a minimum photospheric radius through
eq. (\ref{eq:20}). Combining the limits on both ends, we obtain
\begin{equation}
  \label{eq:6}
  \left( \frac{\pi^4 L_0}{15L_{\rm max}}\right)^{1/6} < R_{\rm ph}/R_0 
  < 1.
\end{equation}
For $L_0/L_{\rm max} = (10^{-2}, 10^{-3}, 10^{-4})$, the
left-hand side of eq. (\ref{eq:6}) $[\pi^4 L_0/(15L_{\rm
  max})]^{1/6} = (0.63, 0.43, 0.29)$ is not far from
unity. Therefore, we conclude that the photospheric radius defined in
eq. (\ref{eq:2}) can be constrained to a fairly narrow range $R_{\rm ph} \in
(R_0/3, R_0)$, where $R_0$ can be determined by one data point
($\nu_0,L_0$) on the R-J tail.

We list in Table \ref{tab:Rph} the constraints on photospheric radii in
some recent TDEs. The jetted TDEs Swift J2058+05, Swift J1644+57 and
some non-jetted TDEs have $R_{\rm ph}\sim$ a few$\times
10^{14}\rm\ cm$.

\label{lastpage}
\end{document}